\documentclass[paper]{JHEP3}

\usepackage{amsmath,amssymb,setspace,graphicx,subfigure}

\JHEPspecialurl{http://jhep.sissa.it/JOURNAL/JHEP3.tar.gz}

\newcommand\fverbdo{\egroup\medskip\noindent%
            \fbox{\unhbox\fverbbox}\ }
\newcommand\fverbit{\egroup\item[\fbox{\unhbox\fverbbox}]}
\newbox\fverbbox

\newcommand{\mpl}{m_{\rm Pl}}
\newcommand{\fnl}{f_{\rm NL}}
\newcommand{\calA}{{\cal A}}
\newcommand{\calL}{{\cal L}}
\newcommand{\calO}{{\cal O}}
\newcommand{\calP}{{\cal P}}

\title{Inflation and dark matter in two Higgs doublet models}
\author{Jinn-Ouk Gong and Hyun Min Lee\\
    Theory Division, CERN, CH-1211 Gev\`eve 23, Switzerland \\
    E-mail: \email{jinn-ouk.gong@cern.ch, hyun.min.lee@cern.ch}}
\author{Sin Kyu Kang \\
Institute of Convergence Fundamental Studies \& School of Liberal Arts \\
Seoul National University of Science and Technology, Seoul 139-931, Republic of Korea \\
 E-mail: \email{skkang@seoultech.ac.kr}}

\preprint{
CERN-PH-TH/2012-024}

\abstract{
We consider the Higgs inflation in the extension of the Standard Model with two Higgs doublets coupled to gravity non-minimally.
In the presence of an approximate global $U(1)$ symmetry in the Higgs sector,  both radial and angular modes of neutral Higgs bosons drive inflation where large non-Gaussianity is possible from appropriate initial conditions on the angular mode. We also discuss the case with single-field inflation for which the $U(1)$ symmetry is broken to a $Z_2$ subgroup. We show that inflationary constraints, perturbativity and stability conditions restrict the parameter space of the Higgs quartic couplings at low energy in both multi- and single-field cases. Focusing on the inert doublet models where $Z_2$ symmetry remains unbroken at low energy, we show that the extra neutral Higgs boson can be a dark matter candidate consistent with the inflationary constraints. The doublet dark matter is always heavy in multi-field inflation while it can be light due to the suppression of the co-annihilation in single-field inflation. The implication of the extra quartic couplings on the vacuum stability bound is also discussed in the light of the recent LHC limits on the Higgs mass.
}

\begin{document}

\section{Introduction}

Cosmic inflation~\cite{inflation} in the very early universe is currently regarded as the leading candidate to resolve a number of cosmological problems, such as the horizon and flatness problems, and to give rise to the initial conditions for the subsequent hot big bang evolution of the universe. Furthermore, one crucial prediction is that during inflation the quantum fluctuations of one or more inflaton fields are stretched to cosmic scales, later becoming the seed of the temperature anisotropies in the cosmic microwave background (CMB) and large scale structure of the universe~\cite{book}. These primordial perturbations have nearly scale invariant power spectrum with almost perfect Gaussian statistics. Most recent observations are consistent with these predictions, strongly supporting inflation in the early universe~\cite{Komatsu:2010fb}.

Implementing inflation in the context of particle physics is, however, a non-trivial task.
The only scalar field in the Standard Model (SM) is the Higgs boson, but it cannot support inflation alone. Thus we usually resort to the theories beyond SM, where generically a number of scalar fields exists.
The merit of the Higgs inflation~\cite{Bezrukov:2007ep,Bezrukov:2010jz} consists in the minimality that we introduce only one additional parameter, the non-minimal coupling of the Higgs doublet to gravity $\xi$~\cite{nonminimalinf}. However, to match the amplitude of the power spectrum we require $\xi = \calO(10^4)$. Such a large value of the non-minimal coupling leads to the problem of unitarity around $\mpl/\xi\sim 10^{13}$ GeV, breaking the perturbative expansion in the vacuum \cite{unitarity}. Therefore, unitarity should be restored by introducing an extra dynamical degree of freedom at the unitarity scale \cite{unitarization}. Furthermore, the recent LHC data~\cite{LHC} set the upper limit on the Higgs mass to $m_h<127\,{\rm GeV}$, which hints that the SM vacuum becomes unstable after renormalization group running at the scale of $10^9-10^{10}$ GeV up to theoretical uncertainties in higher order corrections~\cite{vacuumstability}.
Since the scale of vacuum instability is smaller than the unitarity scale in the Higgs inflation,
the running Higgs quartic coupling would have become already negative even before the non-minimal coupling becomes dominant or unitarity starts being violated.
In this regard, it seems necessary that the original, simplest Higgs inflation should be extended to a non-minimal setup.

In this paper, we consider a slow-roll inflation with non-minimal gravity couplings in the extension of SM with two Higgs doublets\footnote{See Ref.~\cite{review} for review.}. The pseudo-scalar boson of the additional Higgs doublet and the modulus of the Higgs boson lead to multi-field inflation, in the limit of an approximate $U(1)_H$ symmetry in the extended Higgs potential. The Higgs modulus dominates the inflaton dynamics and it takes a similar inflaton potential to the one in the original Higgs inflation. This multi-Higgs inflation is a concrete realization of the toy model with a complex scalar field that has been proposed by two of the authors~\cite{Gong:2011cd}. The modulus of the Higgs boson dominates the slow-roll dynamics while the pseudo-scalar boson can lead to large non-Gaussianity by making a significant change in the inflaton component during the inflation. On the other hand, when $U(1)_H$ is broken to a $Z_2$ parity, there is no multi-field inflation: the mixture of two Higgs moduli drives single-field inflation as in the Higgs portal inflation~\cite{lebedev,singletinflation}, where a real singlet scalar is added to the SM.

We focus on the inert doublet models \cite{inertdm} to discuss the consequences of the inflationary conditions to the Higgs physics at low energy.
First, in the multi-Higgs inflation, the VEVs of two Higgs doublets must be non-zero to generate the potential of the second inflaton, the pseudo-scalar boson,  so there is the possibility that the effective quartic coupling of the Higgs modulus can be small for the Minimal Supersymmetric Standard Model (MSSM)-like boundary conditions on the Higgs quartic couplings during inflation.
This allows for $\calO(1)$ non-minimal couplings without conflicting the Higgs mass at low energy such that there is no unitarity violation below the Planck scale.
If the effective inflaton quartic coupling is of $\calO(1)$, large non-minimal couplings are necessary so unitarity is violated below the Planck scale as in the original Higgs inflation. In this case, we should rely on the UV completion of the two Higgs doublet models as in the Higgs portal inflation \cite{unitarization,lebedev}.
Second, the single-Higgs inflation
may be driven by either pure SM Higgs, pure inert Higgs or mixed Higgs.
The stability conditions on the orthogonal direction to the inflaton restrict the parameter space of the Higgs quartic couplings.
Furthermore, in both multi-Higgs and single-Higgs inflation cases, there is a parameter space for the extra Higgs quartic couplings that is consistent with both the inflationary conditions and the recent LHC limit on the Higgs mass.

We also discuss the low energy constraints on the inert doublet models together with the inflationary ones. In these models, the lightest neutral scalar of the second Higgs doublet can be a dark matter candidate~\cite{inertdm}. In the multi-Higgs inflation, due to an approximate $U(1)_H$ symmetry, there is a small mass splitting between dark matter Higgs and pseudo-scalar, resulting in too large co-annihilation through gauge interactions. Thus, in this case, only heavy dark matter with mass around $600\,{\rm GeV}$ is possible.
On the other hand, in the single-Higgs inflation, the $U(1)_H$ symmetry is broken so the co-annihilation channel is suppressed. Thus, in this case, dark matter of mass smaller than $100\,{\rm GeV}$ is possible through the extra quartic couplings between two Higgs doublets.
Focusing on the pure SM Higgs inflation, we show how the inflationary constraints further reduce the parameter space allowed by collider and dark matter constraints.

The paper is organized as follows. In Section~\ref{sec:2HDM}, we provide the explicit setup of the two Higgs doublet model for inflation. In Section~\ref{sec:inflation}, based on the setup built in the previous section, we address the inflationary dynamics and the resulting constraints on the model parameters. Very interestingly, depending on the value of the Higgs quartic couplings, inflation may be driven by one or more inflaton fields. In Section~\ref{sec:lowenergy}, we discuss the low energy phenomenology and present further constraints on the parameter space. Especially, we show that the inflationary constraints reduce significantly the parameter space consistent with collider and electroweak precision test  data. Finally we conclude in Section~\ref{sec:conc}. Some technical details are relegated to appendices.

\section{Two Higgs doublet model for inflation}
\label{sec:2HDM}

The Jordan frame action for the general two Higgs doublet model is the following,
\begin{equation}
\frac{\calL_J}{\sqrt{-g_J}} = \frac{R}{2}+\Big(\xi_1|\Phi_1|^2+\xi_2|\Phi_2|^2+\xi_3\Phi^\dagger_1 \Phi_2+{\rm c.c.}\Big)R
-\left|D_\mu\Phi_1\right|^2-\left|D_\mu\Phi_2\right|^2-V_J\left(\Phi_1,\Phi_2\right) \, ,
\end{equation}
where the general renormalizable potential is~\cite{review,sm2hd}
\begin{align}
V_J(\Phi_1,\Phi_2) = & \mu^2_1|\Phi_1|^2+\mu^2_2|\Phi_2|^2-\Big(\mu^2_3\Phi^\dagger_1\Phi_2+{\rm c.c.}\Big)
\nonumber\\
& +\frac{1}{2}\lambda_1|\Phi_1|^4+\frac{1}{2}\lambda_2|\Phi_2|^4+\lambda_3 |\Phi_1|^2|\Phi_2|^2+\lambda_4 \left(\Phi^\dagger_1\Phi_2\right)\left(\Phi^\dagger_2\Phi_1\right)
\nonumber\\
&+\left[ \frac{1}{2}\lambda_5\left(\Phi^\dagger_1\Phi_2\right)^2+\lambda_6\left(\Phi^\dagger_1\Phi_1\right) \left(\Phi^\dagger_1\Phi_2\right) + \lambda_7 \left(\Phi^\dagger_2\Phi_2\right) \left(\Phi^\dagger_1\Phi_2\right)+{\rm c.c.} \right] \, .
\end{align}
Here, we have set $\mpl=1$ but it will be restored whenever necessary. As compared to the minimally coupled two Higgs doublet models, we have introduced the non-minimal couplings, $\xi_i$ ($i=1,2,3$), that are assumed to be all positive to avoid a potential instability at large field values.
When the non-minimal coupling $\xi_3$, the mass parameter $\mu_3$ and the quartic couplings $\lambda_6$ and $\lambda_7$ are zero, there is a $Z_2$ symmetry under which the Higgs doublets transform as $\Phi_1\rightarrow \Phi_1$ and $\Phi_2\rightarrow -\Phi_2$ while the SM fermions are neutral.
If this $Z_2$ symmetry is exact, the Higgs doublet $\Phi_2$ would not couple to the SM fermions so that there is no additional flavor violation.
If $\lambda_5=0$ on top of $\mu_3=\lambda_6=\lambda_7=0$,  the symmetry is enhanced to the $U(1)_H$ symmetry.
Henceforth we take a simple choice of dimensionless parameters as $\xi_3=0$ and $\lambda_6=\lambda_7=0$ by imposing the $Z_2$ symmetry,
and assume all the parameters to be real.

First, by making a Weyl transformation of the metric $g^J_{\mu\nu}=g^E_{\mu\nu}/\Omega^2$ with
\begin{equation}\label{conformalxform}
\Omega^2\equiv 1+2\xi_1|\Phi_1|^2+2\xi_2|\Phi_2|^2 \, ,
\end{equation}
we obtain the Einstein frame action as
\begin{align}
\label{einsteinaction}
\frac{\calL_E}{\sqrt{-g_E}} = & \frac{R}{2}-\frac{3}{4}\Big[\partial_\mu \log \left(1+2\xi_1|\Phi_1|^2+2\xi_2|\Phi_2|^2\right)\Big]^2
-\frac{|\partial_\mu\Phi_1|^2+|\partial_\mu\Phi_2|^2}{1+2\xi_1|\Phi_1|^2+2\xi_2|\Phi_2|^2} - V_E(\Phi_1,\Phi_2) \, ,
\\
\label{einsteinpotential}
V_E(\Phi_1,\Phi_2) = & \frac{V_J}{\left(1+2\xi_1|\Phi_1|^2+2\xi_2|\Phi_2|^2\right)^2} \, .
\end{align}
Here we have dropped the gauge interactions.

To discuss the inflationary dynamics,
we take the solutions for two Higgs doublets as
\begin{equation}
\Phi_1=\frac{1}{\sqrt{2}}\left(\begin{array}{l} 0 \\ h_1 \end{array} \right) \, , \quad
\Phi_2=\frac{1}{\sqrt{2}}\left(\begin{array}{l} 0 \\ h_2 e^{i\vartheta} \end{array} \right) \, .  \label{higgsol}
\end{equation}
Then, (\ref{einsteinaction}) and (\ref{einsteinpotential}) become
\begin{align}
\label{einsteinaction2}
\frac{\calL_E}{\sqrt{-g_E}} = & \frac{R}{2}-\frac{3}{4}\Big[\partial_\mu \log\left(1+\xi_1h^2_1+\xi_2h^2_2\right)\Big]^2
-\frac{(\partial_\mu h_1)^2+(\partial_\mu h_2)^2+h^2_2(\partial_\mu\vartheta)^2}{2(1+\xi_1 h^2_1+\xi_2h^2_2)}
-V_E(h_1,h_2,\vartheta) \, ,
\\
\label{einsteinpotential2}
V_E(h_1,h_2,\vartheta) = & (1+\xi_1h^2_1+\xi_2h^2_2)^{-2}\bigg[\frac{1}{2}\mu^2_1 h^2_1+\frac{1}{2}\mu^2_2h^2_2
-\mu^2_3h_1h_2\cos\vartheta
\nonumber \\
& \hspace{3.2cm} +\frac{1}{8}\lambda_1 h^4_1+\frac{1}{8}\lambda_2h^4_2+\frac{1}{4}(\lambda_3 +\lambda_4) h^2_1 h^2_2 +\frac{1}{4}\lambda_5 h^2_1 h^2_2 \cos(2\vartheta)\bigg] \, .
\end{align}
The unbounded from below conditions are
\begin{equation}\label{ufb}
\lambda_1>0\, , \quad \lambda_2>0 \, , \quad \lambda_3+\lambda_4+\lambda_5+\sqrt{\lambda_1\lambda_2}>0 \, .
\end{equation}

Ignoring the mass terms in the potential,
(\ref{einsteinaction2}) and (\ref{einsteinpotential2}) with $\phi^I=\{h_1,h_2,\vartheta\}$ are rewritten as
\begin{equation}
\frac{{\cal L}_E}{\sqrt{-g_E}}= \frac{R}{2}-\frac{1}{2}G_{IJ}\partial_\mu \phi^I\partial^\mu\phi^J-V_E(\phi^I) \, ,
\end{equation}
where
\begin{align}
G_{IJ} = & \frac{1}{1+\xi_1h^2_1+\xi_2 h^2_2}
\begin{pmatrix}
1+\dfrac{6\xi^2_1 h^2_1}{1+\xi_1h^2_1+\xi_2 h^2_2} & \dfrac{6\xi_1\xi_2h_1h_2}{1+\xi_1h^2_1+\xi_2 h^2_2} & 0
\\
\dfrac{6\xi_1\xi_2h_1h_2}{1+\xi_1h^2_1+\xi_2 h^2_2} & 1+\dfrac{6\xi^2_2 h^2_2}{1+\xi_1h^2_1+\xi_2 h^2_2}  & 0
\\
0 & 0 & h^2_2
\end{pmatrix} \, ,
\\
V_E(\phi^I) = & \frac{\lambda_1h^4_1+\lambda_2 h^4_2+2(\lambda_3+\lambda_4)h^2_1h^2_2+2\lambda_5 h^2_1h^2_2\cos(2\vartheta)}{8\left(1+\xi_1h^2_1+\xi_2 h^2_2 \right)^2} \, .
\label{epot2}
\end{align}

Now, taking a large field limit $\xi_1 h^2_1+\xi_2 h^2_2\gg 1$ and making the following field redefinitions,
\begin{align}
\varphi= & \sqrt{\frac{3}{2}}\log(1+\xi_1 h^2_1+\xi_2 h^2_2) \, ,  \label{newfield1}
\\
r = & \frac{h_2}{h_1} \, ,  \label{newfield2}
\end{align}
we find the action in the form~\cite{lebedev}
\begin{align}
\label{largefieldaction}
\frac{\calL_E}{\sqrt{-g_E}} \approx & \frac{R}{2} - \frac{1}{2}\left(1+\frac{1}{6}\frac{r^2+1}{\xi_2 r^2+\xi_1}\right)(\partial_\mu \varphi)^2-\frac{1}{\sqrt{6}}\frac{(\xi_1-\xi_2)r}{\left(\xi_2 r^2+\xi_1\right)^2}(\partial_\mu \varphi)(\partial^\mu r)
\nonumber\\
&-\frac{1}{2}\frac{\xi^2_2 r^2+\xi^2_1}{\left(\xi_2 r^2+\xi_1\right)^3}(\partial_\mu r)^2-\frac{1}{2}\frac{r^2}{\xi_2r^2+\xi_1}\left(1-e^{-2\varphi/\sqrt{6}}\right)(\partial_\mu\vartheta)^2-V_E(\varphi,r,\vartheta) \, ,
\\
\label{potentialwomin}
V_E(\varphi,r,\vartheta)= & \frac{\lambda_1+\lambda_2 r^4+2\lambda_L r^2+2\lambda_5 r^2\cos(2\vartheta)}{8\left(\xi_2 r^2+\xi_1\right)^2}\,\left(1-e^{-2\varphi/\sqrt{6}}\right)^2  \, ,
\end{align}
with $\lambda_L\equiv \lambda_3+\lambda_4$.
The stabilization of the Higgs ratio $r$ will be considered later depending on whether the pseudo-scalar potential is smaller than that of $r$. From now on we omit the subscript $E$.

\section{Inflation driven by the Higgs fields}
\label{sec:inflation}

We divide our discussion into two parts for the multi-Higgs and single-Higgs inflation, depending on the size of the $U(1)_H$ breaking $\lambda_5$ coupling. We analyze how the inflationary conditions restrict the parameter space of the Higgs quartic couplings at low energy in each type of inflation.

\subsection{Multi-Higgs inflation}

Suppose that $\lambda_5\ll \lambda_1$, $\lambda_2$, $\lambda_3$, $\lambda_4$, which is the case with an approximate $U(1)_H$ symmetry. In this case, the potential term for $\vartheta$ does not affect the stabilization of the orthogonal mode $r$ much but it gives rise to a small tilt for both $\varphi$ and $\vartheta$. Then, the part of the potential independent of $\varphi$ and $\vartheta$ becomes
\begin{equation}\label{rpot}
V_{\varphi,\vartheta\text{-indep}} \approx \frac{\lambda_1+\lambda_2 r^4+2\lambda_L r^2}{8\left(\xi_1+\xi_2 r^2\right)^2} \, .
\end{equation}
After stabilizing $r$ at the minimum $r_0$ as in Appendix~\ref{app:stabilization}, from (\ref{epot2}), we find the potential for one of the neutral Higgses and the pseudo-scalar Higgs as
\begin{equation}\label{effpot}
V(\varphi,\vartheta) \approx \frac{\lambda_{\rm eff}}{4\xi^2_{\rm eff}} \left(1-e^{-2\varphi\sqrt{6}}\right)^2 \left[1+\delta \cos(2\vartheta)\right] \, ,
\end{equation}
where
$\delta \equiv\lambda_5r^2_0/\lambda_{\rm eff}$,
$\xi_{\rm eff}\equiv\xi_1+\xi_2 r^2_0$ and
$\lambda_{\rm eff}\equiv\left(\lambda_1+\lambda_2 r^4_0+2\lambda_L r^2_0\right)/2$,
with the finite value of $r^2_0$ given by
\begin{equation}\label{finiter0}
r^2_0=\frac{\lambda_1\xi_2-\lambda_L\xi_1}{\lambda_2\xi_1-\lambda_L\xi_2} \, .
\end{equation}
In this case, the effective non-minimal coupling and the effective quartic coupling are
\begin{align}
\lambda_{\rm eff} = & \frac{\lambda_1\lambda_2-\lambda^2_L}{2} \frac{\lambda_1\xi^2_2+\lambda_2\xi^2_1-2\lambda_L\xi_1\xi_2}{\left(\lambda_2\xi_1-\lambda_L\xi_2\right)^2} \, ,
\\
\xi_{\rm eff} = & \frac{\lambda_1\xi^2_2+\lambda_2\xi^2_1-2\lambda_L\xi_1\xi_2}{\lambda_2\xi_1-\lambda_L\xi_2} \, .
\end{align}
Then, the inflationary vacuum energy becomes
\begin{equation}
V_0=\frac{\lambda_1\lambda_2-\lambda^2_L}{8\left(\lambda_1\xi^2_2+\lambda_2\xi^2_1-2\lambda_L\xi_1\xi_2\right)} \, .
\end{equation}
For the above minimum with finite $r_0$ to be present, we need to impose the following conditions,
\begin{align}
\lambda_1\xi_2-\lambda_L\xi_1>&0 \, ,
\label{cond1}\\
\lambda_2\xi_1-\lambda_L\xi_2>& 0 \, ,
\label{cond2} \\
\lambda_1\lambda_2-\lambda_L^2>&0 \, .
\label{cond3}
\end{align}
The last condition is required for the absence of deep minima which make the electroweak vacuum metastable and for a positive vacuum energy during inflation as well. We note that if (\ref{cond1}) and (\ref{cond2}) are not satisfied, $r_0=0$ or $r_0=\infty$, so a single neutral Higgs boson drives inflation.

Here, we find that since $\lambda_1>0$ and $\lambda_2>0$, for $\lambda_L>0$, the third condition (\ref{cond3}) comes out automatically. In this case, in particular for the inert doublet model, the first condition (\ref{cond1}) becomes stronger than the vacuum stability bound on the SM Higgs quartic coupling.
Then, the unbounded from below conditions (\ref{ufb}) give no further constraint.
On the other hand, for $\lambda_L<0$ , (\ref{cond1}) and (\ref{cond2}) are trivially satisfied and the third condition (\ref{cond3}) is the only constraint.
Thus, for a small $\lambda_5$, the third condition is approximately the same as the unbounded from below conditions (\ref{ufb}).
That is, (\ref{ufb}) guarantees the positive vacuum energy during inflation.
We note that in order to make the SM Higgs boson below the recent LHC limits ($m_h<127\,{\rm GeV}$) compatible with vacuum stability, $\lambda_L<0$ is preferred. This is different from the single-Higgs inflation, as we will see shortly.

For most of the parameter space, $\lambda_{\rm eff}$ is not small so the CMB normalization of density perturbations requires $\xi_{\rm eff}$ to be of $\calO(10^4)$ as will be shown in next sections. Therefore, unitarity is violated at $\mu_U\sim \mpl/\xi_{\rm eff}$, which is much below the Planck scale. But, it is possible to maintain the inflationary conditions (\ref{cond1}), (\ref{cond2}) and (\ref{cond3}) in the unitarization process of introducing a heavy real scalar as in Higgs portal inflation~\cite{unitarization,lebedev}.

We remark that the effective self-coupling of the inflaton, $\lambda_{\rm eff}$, does not have to be necessarily of $\calO(1)$ to satisfy the Higgs mass constraint, unlike the SM Higgs inflation. In particular, the quartic couplings in MSSM are given by $\lambda_1=\lambda_2=\big(g^2+{g'}^2\big)/4$, $\lambda_3=\big(g^2-{g'}^2\big)/4$, $\lambda_4=-g^2/2$ and $\lambda_5=\lambda_6=\lambda_7=0$.
Then, MSSM would lead to $\lambda_1\lambda_2= \lambda^2_L$ such that $\lambda_{\rm eff}=0$. Even for split supersymmetry~\cite{split-susy} in which gauginos and Higgsinos have weak-scale masses while the other superpartners are very heavy, the MSSM relations between the quartic couplings are still RG invariant.
But, when a heavy singlet or an $SU(2)$ triplet
couples to the Higgs doublets, the threshold correction could lead to a small deviation from the MSSM relations between the quartic couplings~\cite{giudice}.
On the other hand, in the NMSSM, a light SM singlet leads to a deviation in the $\lambda_3$ coupling from the MSSM value such that the Higgs potential becomes positive at large Higgs values with $\tan\beta=1$~\cite{susyinflation}.
However, in the supersymmetric models for the non-minimal gravity couplings, the $U(1)_H$ preserving couplings, $\xi_1$ and $\xi_2$, are fixed to $-1/6$, because they are related to the Higgs kinetic terms in Jordan frame.
Instead, the $U(1)_H$-breaking coupling $\xi_3$ appears as a (anti-)holomorphic term in Jordan frame supergravity and it can be arbitrary.  Therefore, there is no counterpart of our multi-field inflation with large positive non-minimal couplings, $\xi_1$ and $\xi_2$, in the supersymmetric models.

In multi-Higgs inflation, the effective action of the canonical inflaton fields becomes, at large field values $\xi_1|h_1|\gg1$ and $\xi_2|h_2|\gg1$,
\begin{equation}\label{multiHiggsaction}
S = \int d^4x \sqrt{-g} \left[ \frac{R}{2} - \frac{1}{2}(\partial_\mu\varphi)^2 - \frac{1}{2}e^{2b(\varphi)}(\partial_\mu\chi)^2 - W(\varphi,\chi) \right] \, ,
\end{equation}
where  $e^{2b(\varphi)} \equiv 1-e^{-2\varphi/\sqrt{6}}$. Here, the potential is of the product form $W(\varphi,\chi) = U(\varphi)V(\chi)$ with
\begin{align}
U(\varphi) = & \frac{\lambda}{4\xi^2} \left( 1 - e^{-2\varphi/\sqrt{6}} \right)^2 \, ,
\\
V(\chi) = & 1 + \delta\cos \left( 2\sqrt{\xi}\chi \right) \, ,
\end{align}
where $\chi\equiv \vartheta/\sqrt{\xi}$, $\xi\equiv\xi_\text{eff}/r_0^2$ and $\lambda\equiv\lambda_\text{eff}/r_0^4$. So, in terms of the effective couplings, $\lambda$, $\xi$ and $\delta$, (\ref{multiHiggsaction}) in two Higgs doublet models coincides with the one in a toy model for inflation with a complex scalar field in which a small violation of the $U(1)$ global symmetry was taken~\cite{Gong:2011cd}.

In our model, we obtain the slow-roll parameters as $\epsilon =  \epsilon^\varphi + \epsilon^\chi $, where
\begin{align}
\epsilon^\varphi =& \frac{4}{3}\, \frac{e^{-4\varphi/\sqrt{6}}}{\left(1-e^{-2\varphi/\sqrt{6}}\right)^2} \, ,
\\
\epsilon^\chi =& \frac{1}{1-e^{-2\varphi/\sqrt{6}}} \frac{2\xi \delta^2\sin^2\left(2\sqrt{\xi}\chi\right)}{\left[1+\delta \cos\left(2\sqrt{\xi}\chi\right)\right]^2} \, ,
\end{align}
and
\begin{align}
\eta^{\varphi\varphi} = & -\frac{4}{3}e^{-2\varphi/\sqrt{6}} \,\frac{1-2e^{-2\varphi/\sqrt{6}}}{\left(1-e^{-2\varphi/\sqrt{6}}\right)^2} \, ,
\\
\eta^{\chi\chi} = &-\frac{1}{1-e^{-2\varphi/\sqrt{6}}} \frac{4\xi\delta\cos\left(2\sqrt{\xi}\chi\right)}{1+\delta\cos\left(2\sqrt{\xi}\chi\right)} \, .
\end{align}
Then, the radial mode $\varphi$ dominates the slow-roll condition and ends inflation as in the SM Higgs inflation, while the angular mode $\chi$ keeps slow-rolling and takes a sub-dominant fraction in the slow-roll parameter $\epsilon$.
Since the number of $e$-folds $N$ for a product potential is given by
\begin{equation}\label{efold0}
N = \int_e^\star \frac{U}{U'} d\varphi = \int_e^\star\, e^{2b}\frac{V}{V'} \, d\chi \, ,
\end{equation}
we obtain
\begin{equation}\label{efold}
N = \frac{3}{4} \left[ e^{2\varphi_\star/\sqrt{6}} - e^{2\varphi_e/\sqrt{6}} - \frac{2}{\sqrt{6}} \left( \varphi_\star - \varphi_e \right) \right] \, ,
\end{equation}
where the subscripts $\star$ and $e$ respectively denote the moment when the scale of our interest exits the horizon and the end of slow-roll inflation.
The slow-roll condition is violated {\em mainly} by $\varphi$ when $e^{-2\varphi_e/\sqrt{6}} \approx 0.464$, so, for $N = 60$, we need $e^{2\varphi_\star/\sqrt{6}} \approx 80.5$. On the other hand, from the second equality in (\ref{efold0}), the background evolution of $\chi$ is fixed by
\begin{equation}
\left|\tan\left(\sqrt{\xi}\chi_e\right)\right| \approx \left|\tan\left(\sqrt{\xi}\chi_\star\right)\right| e^{4\xi\delta N} \, .
\end{equation}
Thus, we can determine the final fraction for a given initial fraction in the $\epsilon$ parameter. Defining
\begin{align}
\cos^2\theta \equiv & \frac{\epsilon^\varphi}{\epsilon} \, ,
\\
\sin^2\theta \equiv & \frac{\epsilon^\chi}{\epsilon} \, ,
\end{align}
we can find
\begin{equation}
\theta_e \sim \sqrt{\epsilon_\star} e^{4\xi\delta N} \theta_\star \, .
\end{equation}
Depending on the values of $\xi$ and $\delta$, we can naturally have $\theta_e = \calO(100)\theta_\star$.

\subsubsection{Constraints from multi-Higgs inflation}

Now we consider the constraints on the dimensionless parameters of two Higgs doublets coming from the inflationary dynamics, focusing on the multi-Higgs inflation.

Unitarity in the vacuum would be violated at the scale $\mu_U={\rm min}(1/\xi_1,1/\xi_2)$, lower than the Planck scale, for large non-minimal couplings of the Higgs doublets.
Thus, we impose inflationary conditions,  perturbativity and stability at unitarity scale $\mu_U$. A UV completion of the model may keep the constraints unchanged \cite{lebedev}, apart from that we imposed those conditions at the scale a bit lower than the scale of inflation, $\calO(1/\xi_{\rm eff})$.
Perturbativity and stability conditions for the Higgs potential at the unitarity scale are
\begin{align}\label{pert}
|\lambda_i| < & \pi \, ,
\\
\label{stable}
\lambda_1, \, \lambda_2 > & 0 \, .
\end{align}
The additional condition for vacuum stability as shown in (\ref{ufb}) is imposed as well.  These conditions are to be satisfied in the single-Higgs inflation too.

We apply the constraints from the observations on the scalar power spectrum $\calP_\zeta$, its index $n_\zeta$ and the non-linear parameter $\fnl$. We can compute these quantities using the $\delta{N}$ formalism~\cite{deltaN}, which is conformally invariant~\cite{Gong:2011qe}. They are given by
\begin{align}
\label{powerspectrum}
\calP_\zeta = & \left( \frac{H_\star}{2\pi} \right)^2 \frac{1}{2\epsilon_\star} e^{2X}\frac{\cos^4\theta_e}{\sin^2\theta_\star} \left( \calA^2\tan^2\theta_\star + \tan^4\theta_e \right) \, ,
\\
\label{spectralindex}
n_\zeta - 1 = & -2\epsilon_\star - 4e^{-2X}\frac{\sin^2\theta_\star}{\cos^4\theta_e \left( \calA^2\tan^2\theta_\star + \tan^4\theta_e \right)}\epsilon_\star
\nonumber\\
& + \frac{\cos^2\theta_\star}{12} \frac{\left( \calA\tan^2\theta_\star - \tan^2\theta_e \right)^2}{\calA^2\tan^2\theta_\star + \tan^4\theta_e} \left( \eta^b_\star + 2\epsilon^b_\star \right) \epsilon_\star + \frac{8\calA\sin^2\theta_\star\tan^2\theta_e}{\calA^2\tan^2\theta_\star + \tan^4\theta_e}\epsilon_\star
\nonumber\\
& - \cos^2\theta_\star\tan^2\theta_e \frac{2\calA\tan^2\theta_\star - \tan^2\theta_e}{\calA^2\tan^2\theta_\star + \tan^4\theta_e}\epsilon_\star + \frac{2\left( \calA^2\tan^2\theta_\star\eta^{\varphi\varphi}_\star + \tan^4\theta_e\eta^{\chi\chi}_\star \right)}{\calA^2\tan^2\theta_\star + \tan^4\theta_e} \, ,
\end{align}
\begin{align}\label{nG}
\frac{6}{5}\fnl = & \frac{e^{-X}}{\left( \calA^2\tan^2\theta_\star + \tan^4\theta_e \right)^2} \left( -\frac{\calA^3\tan^4\theta_\star}{\cos^2\theta_e}\eta^{\varphi\varphi}_\star - \frac{\tan^6\theta_e}{\cos^2\theta_e}\eta^{\chi\chi}_\star \right)
\nonumber\\
& + 2e^{-X}\frac{\sin^2\theta_\star}{\cos^2\theta_e} \frac{\calA^3\tan^2\theta_\star + \tan^6\theta_e}{\left( \calA^2\tan^2\theta_\star + \tan^4\theta_e \right)^2}\epsilon_\star + e^{-X} \frac{\sin^2\theta_\star}{\cos^2\theta_e} \frac{\calA^2\tan^2\theta_e\tan^2\theta_\star}{\left( \calA^2\tan^2\theta_\star + \tan^4\theta_e \right)^2}\epsilon_\star
\nonumber\\
& + 2\tan^2\theta_e \frac{\left( \calA\tan^2\theta_\star - \tan^2\theta_e \right)^2}{\left( \calA^2\tan^2\theta_\star + \tan^4\theta_e \right)^2} \left\{ \eta^{\chi\chi}_e\cos^2\theta_e + \sin^2\theta_e \left[ \eta^{\varphi\varphi}_e - \epsilon_e \left( 4\cos^2\theta_e + \frac{1}{2}\sin^2\theta_e \right) \right] \right\} \, ,
\end{align}
where $X\equiv 2b_e-2b_*$, $\epsilon^b = \epsilon^\varphi$, $\eta^b = -4e^{2\varphi/\sqrt{6}}\epsilon^\varphi$ and
$\calA \equiv e^{-X} \left[ 1 + \left( 1-e^X \right) \tan^2\theta_e \right]$.

According to the most recent WMAP7 observations~\cite{Komatsu:2010fb}, $\calP_\zeta$, $n_\zeta$ and $\fnl$ are constrained by
\begin{align}
\calP_\zeta = \, & \left( 2.430 \pm 0.091 \right) \times 10^{-9} \, ,
\\
n_\zeta = \, & 0.968 \pm 0.012 \, ,
\\
-10 < & ~\fnl < 74 \, ,
\end{align}
respectively. Here, the quoted errors are $1\sigma$ for $\calP_\zeta$ and $n_\zeta$, while $2\sigma$ for $\fnl$. Among these observations, from $V \approx \lambda/\left(4\xi^2\right)$, the constraint on $\calP_\zeta$ essentially enables us to replace $\xi$ with $\lambda$. This gives, using the central value of $\calP_\zeta$,
\begin{equation}\label{xieff-lambdaeff}
\xi = \frac{\sqrt{\lambda}}{\sqrt{24\pi^2 \times 4\epsilon_\star \times 2.430 \times 10^{-9}}} \sim 5\times10^4\sqrt{\lambda} \, .
\end{equation}
Then, we are left with $\lambda$ and $\delta$, or $\lambda_\text{eff}$ and $\lambda_5$ which are more directly relevant for the Higgs. From now on we work with $\lambda_\text{eff}$ and $\lambda_5$.

\begin{figure}[!t]
 \begin{center}
  \includegraphics[width=15cm]{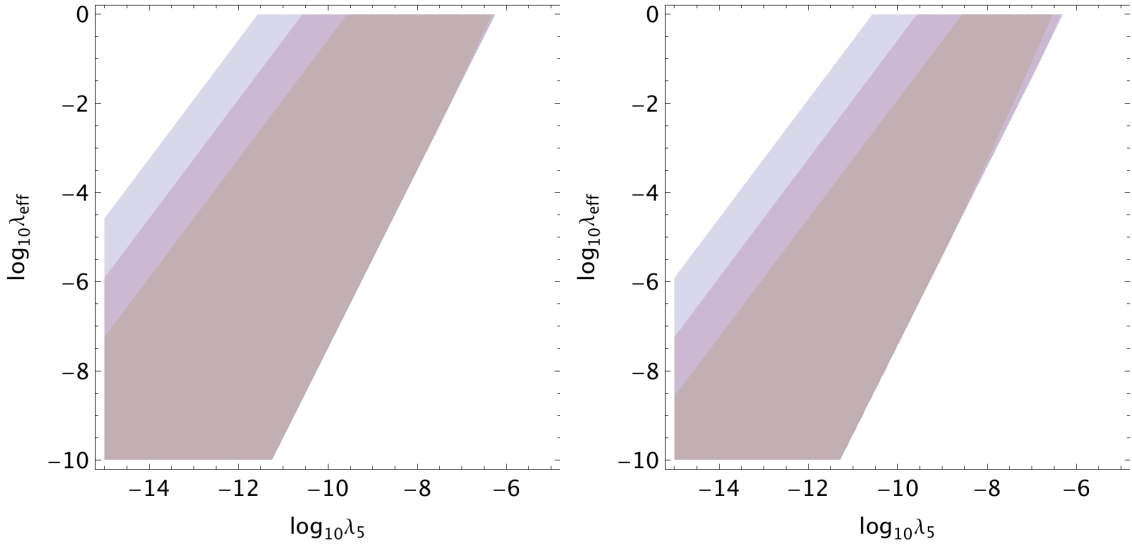}
 \end{center}
\caption{Allowed parameter region of $\lambda_\text{eff}$ and $\lambda_5$. We have considered two cases, (left) $\theta_\star = 10^{-5}$ and (right) $\theta_\star = 10^{-4}$. We have also chosen different values of $r_0$: From the top to bottom, $r_0=1$, 10 and 100, respectively. We can see the tendency that the closer to the hilltop $\chi$ starts initially, the larger we have parameter region consistent with observations.}
 \label{fig:parameterregion}
\end{figure}

Now we turn to the constraints from $n_\zeta$ and $\fnl$. In Figure~\ref{fig:parameterregion} we show the allowed region of $\lambda_\text{eff}$ and $\lambda_5$ from the observations on $n_\zeta$ and $\fnl$. We can understand this qualitatively by considering the simple case\footnote{In this case, large non-Gaussianity is possible. See Ref.~\cite{Gong:2011cd} for a more complete analysis on the conditions for large non-Gaussianity in the same toy model.} $\calA^2\tan\theta_\star^2 \lesssim \tan^4\theta_e$, where we can approximate
\begin{equation}
\fnl \approx -\frac{10}{3|\delta|} \sqrt{\frac{2\xi\delta^2}{\sin^2\theta_e} \left( \frac{2\xi\delta^2}{\sin^2\theta_e-1} \right)} \, .
\end{equation}
Then, with the coefficients of $\eta_\star^{\varphi\varphi}$ and $\eta_\star^{\chi\chi}$ not abruptly small, and with the typical value $\eta_\star^{\varphi\varphi} = \calO(0.01)$, using $\left|\eta_\star^{\chi\chi}\right| \sim \left|\eta_\text{max}^{\chi\chi}\right| \sim 4\xi|\delta|$ from $n_\zeta$ we find
\begin{equation}\label{indexconstraint}
\xi|\delta| \lesssim 0.01 \longleftrightarrow \frac{\lambda_5}{\sqrt{\lambda_\text{eff}}} \lesssim 2\times10^{-7} \, .
\end{equation}
From $\fnl$, essentially we have two constraints. First, we demand that it is real so that
\begin{equation}\label{positivefNLconstraint}
\xi\delta^2 \gtrsim \frac{\sin^2\theta_e}{2} \longleftrightarrow r_0^2 \frac{\lambda_5^2}{\lambda_\text{eff}^{3/2}} \gtrsim 10^{-5} \sin^2\theta_e \, .
\end{equation}
Note that in this case we always have $\fnl <0$. Another constraint is that $|\fnl |\lesssim10$, from which given that the square root gives $\calO(1)$,
\begin{equation}\label{fNLconstraint}
|\fnl| \lesssim 10 \longleftrightarrow \frac{\lambda_5}{\sqrt{\lambda_\text{eff}}} \lesssim 10^{-3} \sin^2\theta_e \, .
\end{equation}
This constraint may be either stronger or weaker than (\ref{indexconstraint}). But too small $\theta_e$ would require more finely tuned initial condition for $\chi_\star$. Combining these constraints (\ref{indexconstraint}), (\ref{positivefNLconstraint}) and (\ref{fNLconstraint}), we can see that the allowed parameter region of $\lambda_\text{eff}$ and $\lambda_5$ is bounded.

For $\xi_{\rm eff}=\calO(1)$, we need $\sqrt{\lambda_{\rm eff}}\sim 10^{-5}$ so that $|\lambda_5|\lesssim 10^{-12}$ if we take (\ref{indexconstraint}). In this case, small $\lambda_{\rm eff}$ can be due to a small deviation from the MSSM boundary condition at high scale, while small $\lambda_5$ can be attributed to small breaking of the $U(1)_H$ symmetry in the Higgs sector. In the other extreme limit $\xi_{\rm eff}=\calO(10^4)$, we need $\lambda_{\rm eff}=\calO(1)$. Further, as seen in Appendix~\ref{app:kinmixing}, we find $r_0\lesssim\sqrt{\xi_{\rm eff}}\sim 100$, for the heavy Higgs modes of the mass squared, $m^2\gtrsim H^2$. Therefore, the effective quartic coupling $\lambda_{\rm eff}$ may lie in the wide range between $\calO\left(10^{-10}\right)$ and $\calO(1)$. On the other hand,  the $U(1)_H$-breaking quartic coupling, $|\lambda_5|$, remains very small for all the range of $\lambda_{\rm eff}$.

\begin{figure}[!t]
\begin{center}
 \includegraphics[width=15cm]{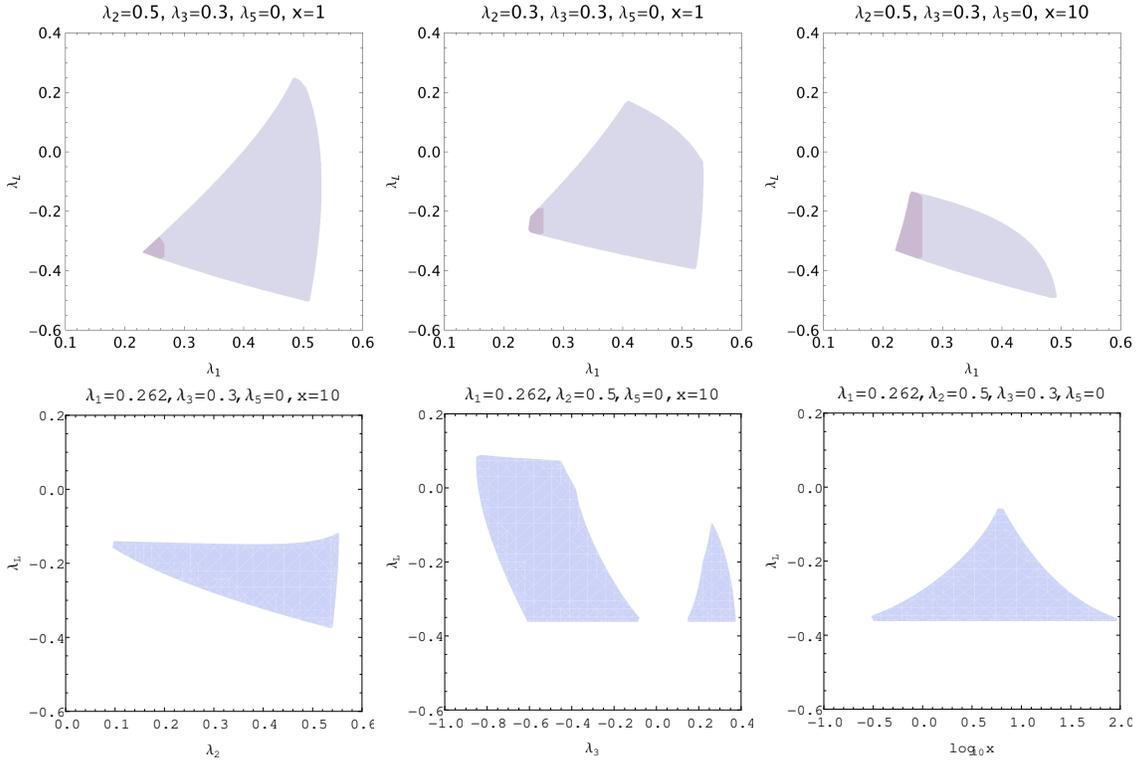}
\end{center}
\caption{Parameter space consistent with multi-Higgs inflation. $\lambda_i$ are given at the scale $m_t$, while $x\equiv \xi_2/\xi_1$ is a high energy
input. Shaded area satisfies the LHC limit on the Higgs mass while the LEP bound on the Higgs mass is automatically satisfied. $m_h=126$ GeV is assumed in the lower row. }
\label{fig:multiHiggsconst}
\end{figure}

\begin{figure}[!t]
 \begin{center}
  \includegraphics[width=10cm]{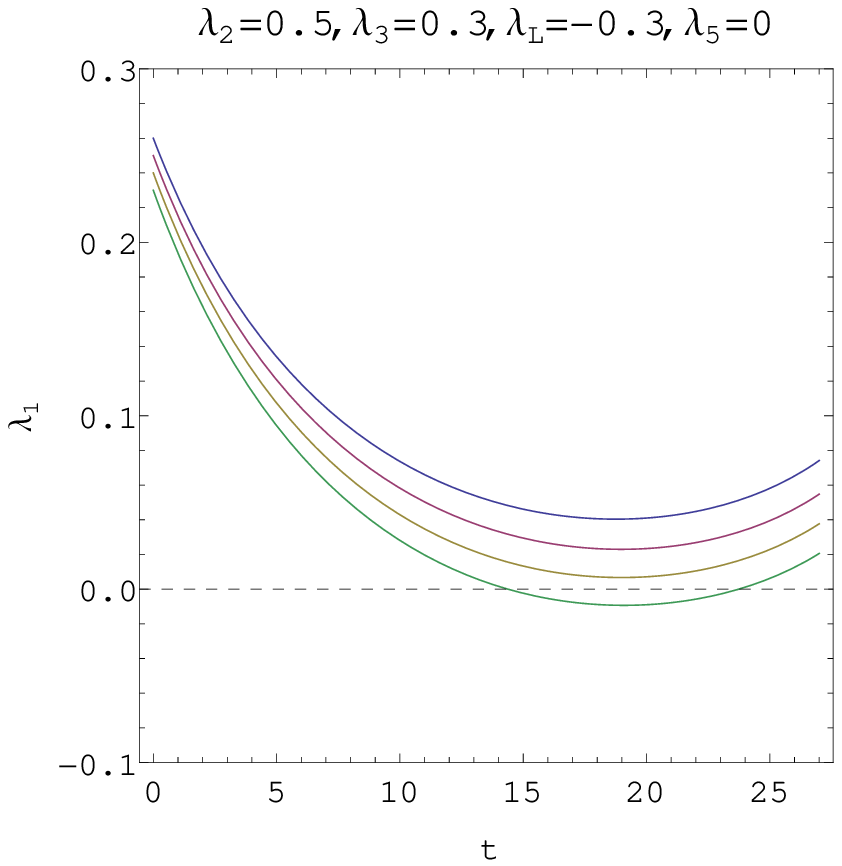}
 \end{center}
\caption{SM Higgs quartic coupling versus $t=\log(\mu/m_W)$ for $\lambda_1=0.26$, 0.25, 0.24 and 0.23 (which corresponds to $m_h=125$, 123, 120 and 118 GeV, respectively) at $\mu=m_t$ from top to bottom. At the unitarity scale, $t_U=\log(\mu_U/m_t)=26$.}
\label{fig:lambda1running}
\end{figure}

In Figure~\ref{fig:multiHiggsconst}, by numerically solving the renormalization group equations for the Higgs quartic couplings given in Appendix~\ref{app:RG} from the unitarity scale~\cite{lebedev,singletinflation,inflationRG}, we show the parameter space of the low energy quartic couplings in the $U(1)_H$ symmetry limit\footnote{We note that $\lambda_5$ should not be exactly zero for dark matter detection as discussed in the later section, but the size of $\lambda_5$ relevant for multi-field inflation is too small to affect the running of the other quartic couplings.}, being consistent with the high scale constraints (\ref{cond1}), (\ref{cond2}), (\ref{cond3}), (\ref{pert}), (\ref{stable}) at the unitarity scale and the unbounded from below constraints (\ref{ufb}) at low energy.
In particular, we find that there is a small parameter space where the vacuum stability is guaranteed until unitarity scale due to $\lambda_3$ and $\lambda_L$ couplings contributing positively to the beta function of the SM Higgs quartic coupling. The parameter space consistent with the Higgs mass bound is small, because the other coupling between the two Higgs bosons, $\lambda_5$, is very small.
The running of the SM Higgs quartic coupling to low energy is shown explicitly in Figure~\ref{fig:lambda1running} for the SM Higgs quartic couplings compatible with the LHC limits on the Higgs mass  in a certain parameter region.

We end this subsection with a remark on the constraint on the $U(1)_H$-breaking dimensionless parameters. If $\xi_3$, $\lambda_6$ and $\lambda_7$ are non-zero, we would have additional terms in the inflaton potential as follows,
\begin{equation}
\delta V(\chi) \approx \left[\frac{2}{\lambda_{\rm eff}} \left(\lambda_6r_0+\lambda_7 r^3_0\right) - \frac{2\xi_3r_0}{\xi_{\rm eff}}\right]\cos \left(\sqrt{\xi}\chi\right) \, .
\end{equation}
Thus, for $n_\zeta$ and $\fnl$ to be consistent with WMAP7, we need
\begin{equation}
\frac{|\xi_3|r_0}{\xi_{\rm eff}}\, , \frac{1}{\lambda_{\rm eff}}\left(\lambda_6r_0+\lambda_7 r^3_0\right)\lesssim 0.01 \frac{r^2_0}{\xi_{\rm eff}} \, .
\end{equation}
Consequently,  for $r_0=\calO(1)$, we find $|\xi_3|\lesssim 0.01$ and additional constraints on $\lambda_6$ and $\lambda_7$ similar to (\ref{indexconstraint}). As discussed earlier, we can set $\xi_3=\lambda_6=\lambda_7=0$ by imposing the $Z_2$ discrete symmetry while $\lambda_5$ survives.
The loop corrections to all the $U(1)_H$ breaking dimensionless couplings are suppressed if small $U(1)_H$ breaking at tree level is imposed. In particular, since $\lambda_5$ is very small during inflation, it remains small at low energy, becoming negligible for Higgs physics.
On the other hand, the mass parameter, $\mu_3$, would lead to a soft breaking of both the $U(1)_H$ symmetry and the $Z_2$ symmetry but it is not constrained by the inflationary constraints.

\subsection{Single-Higgs inflation}

When $\lambda_5$ is not small enough, i.e. $|\lambda_5|> 10^{-7}$, there is no slow-roll along the pseudo-scalar Higgs but rather it is stabilized. In this case, single field inflation is driven by one of the CP-even Higgs bosons so the inflationary conditions are different from those in multi-field inflation.  Moreover, since the $U(1)_H$-breaking $\lambda_5$ coupling is sizable, it can affect the running of the other quartic couplings to low energy.

From the Higgs potential (\ref{potentialwomin}), after $\vartheta$ is stabilized, the $\varphi$-independent part of the potential becomes
\begin{equation}
V_{\varphi\text{-indep}} \approx \frac{\lambda_1+\lambda_2r^4 +2{\widetilde\lambda}_L r^2}{8\left(\xi_1+\xi_2 r^2\right)^2} \, ,
\end{equation}
with ${\widetilde\lambda}_L\equiv\lambda_L-|\lambda_5|$.
After stabilizing the Higgs ratio at the minimum as in the previous section, we find that the  potential for the single-Higgs inflation as
\begin{equation}\label{effpot2}
V \approx \frac{\lambda_{\rm eff}}{4\xi^2_{\rm eff}}\left(1-e^{-2\varphi\sqrt{6}}\right)^2 \, ,
\end{equation}
where
$\xi_{\rm eff}\equiv\xi_1+\xi_2 r^2_0$ and
$\lambda_{\rm eff}\equiv\big(\lambda_1+\lambda_2 r^4_0+2{\widetilde\lambda}_L r^2_0\big)/2$. In this case, the inflationary predictions are the same as those in the original Higgs inflation. That is, for the number of $e$-folds, $N=60$, we obtain the spectral index and the tensor-to-scalar ratio as
\begin{equation}
n_\zeta \approx 0.966 \, , \quad r \approx 3\times 10^{-3} \, .
\end{equation}

\begin{figure}[!t]
 \begin{center}
  \includegraphics[width=15cm]{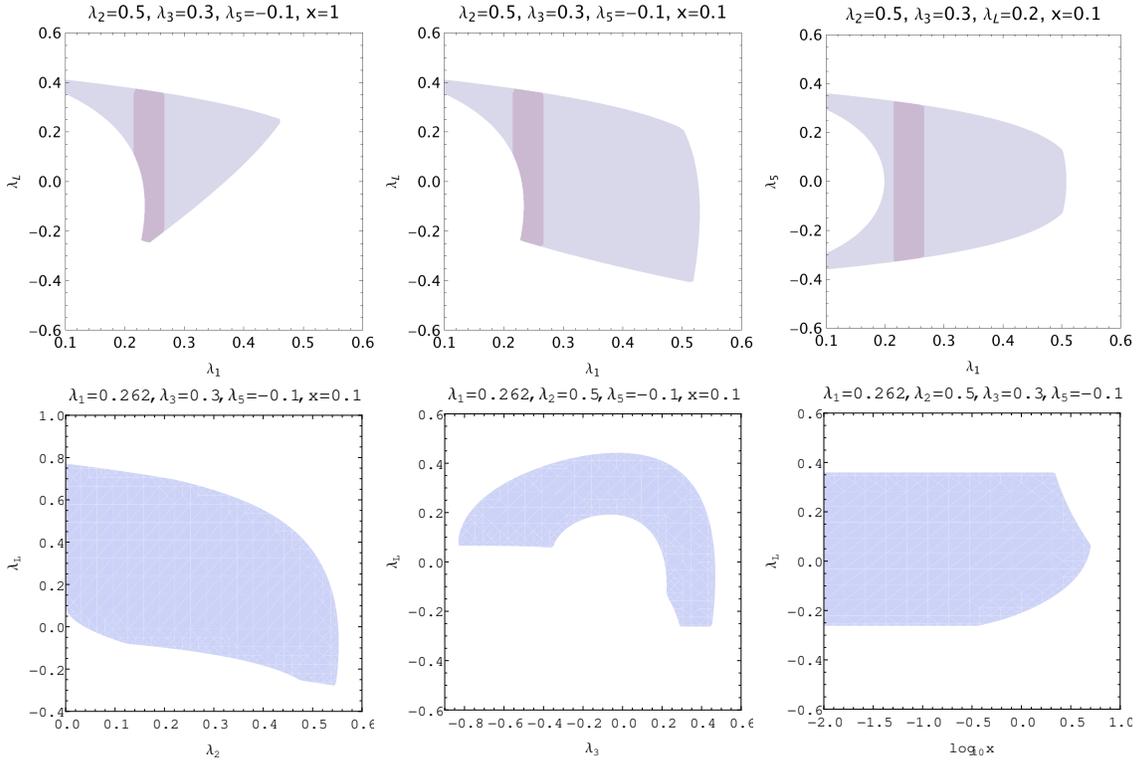}
 \end{center}
\caption{Parameter space consistent with the pure SM Higgs
inflation. Shaded area satisfies both the LEP and LHC limits on the Higgs mass. $m_h=126\,{\rm GeV}$ is assumed in the lower row. }
\label{fig:pureSMHiggsconst}
\end{figure}

\subsubsection{Mixed Higgs inflation}

For a finite $r_0$, the inflaton is a mixture of two CP-even Higgs bosons. The inflationary vacuum that we obtained is similar to the multi-Higgs inflation with the pseudo-scalar boson being frozen. The same formulas (\ref{finiter0})-(\ref{cond3}) with $\lambda_L$ being replaced by ${\widetilde\lambda}_L$ are applied in this case. In the limit of a small $\lambda_5$, the inflationary conditions are the same as in multi-Higgs inflation.  Otherwise, a sizable $\lambda_5$ would shift the allowed parameter space of $\lambda_L=\lambda_3+\lambda_4$ by a positive value.

For large non-minimal couplings, unitarity can be restored while the counterpart of inflationary conditions (\ref{cond1}), (\ref{cond2}) and (\ref{cond3}) with $\lambda_L$ being replaced by ${\widetilde\lambda}_L$ is maintained \cite{lebedev} as in the multi-Higgs inflation.

\subsubsection{Pure Higgs inflation}

For $r_0=0$ or $r_0=\infty$, only one of CP-even Higgs bosons plays the role of the inflaton. From Appendix~\ref{app:stabilization}, we obtain the vacuum energy and the inflationary conditions: for $r_0=0$,
\begin{equation}
V \approx \frac{\lambda_1}{8\xi^2_1} \left(1-e^{-2\varphi/\sqrt{6}}\right)^2~;
\end{equation}
\begin{equation}
\label{oneHcond1}
\lambda_1 \xi_2-{\widetilde\lambda}_L \xi_1 <0 \, , \quad \lambda_2\xi_1-{\widetilde\lambda}_L \xi_2>0~;
\end{equation}
for $r_0=\infty$,
\begin{equation}
V \approx \frac{\lambda_2}{8\xi^2_2} \left(1-e^{-2\varphi/\sqrt{6}}\right)^2~;
\end{equation}
\begin{equation}
\label{oneHcond2}
\lambda_1 \xi_2-{\widetilde\lambda}_L \xi_1 >0 \, , \quad \lambda_2\xi_1-{\widetilde\lambda}_L \xi_2<0 \, .
\end{equation}
We note that as compared to the multi-Higgs inflation, the minimum conditions for the inflationary vacuum for the single-Higgs inflation are different. In particular, due to the first condition in (\ref{oneHcond1}) or the second condition
in (\ref{oneHcond2}), a positive value of ${\widetilde\lambda}_L$ is preferred in the single-Higgs inflation, unlike the multi-Higgs inflation.

\begin{figure}[!t]
\begin{center}
 \includegraphics[width=15cm]{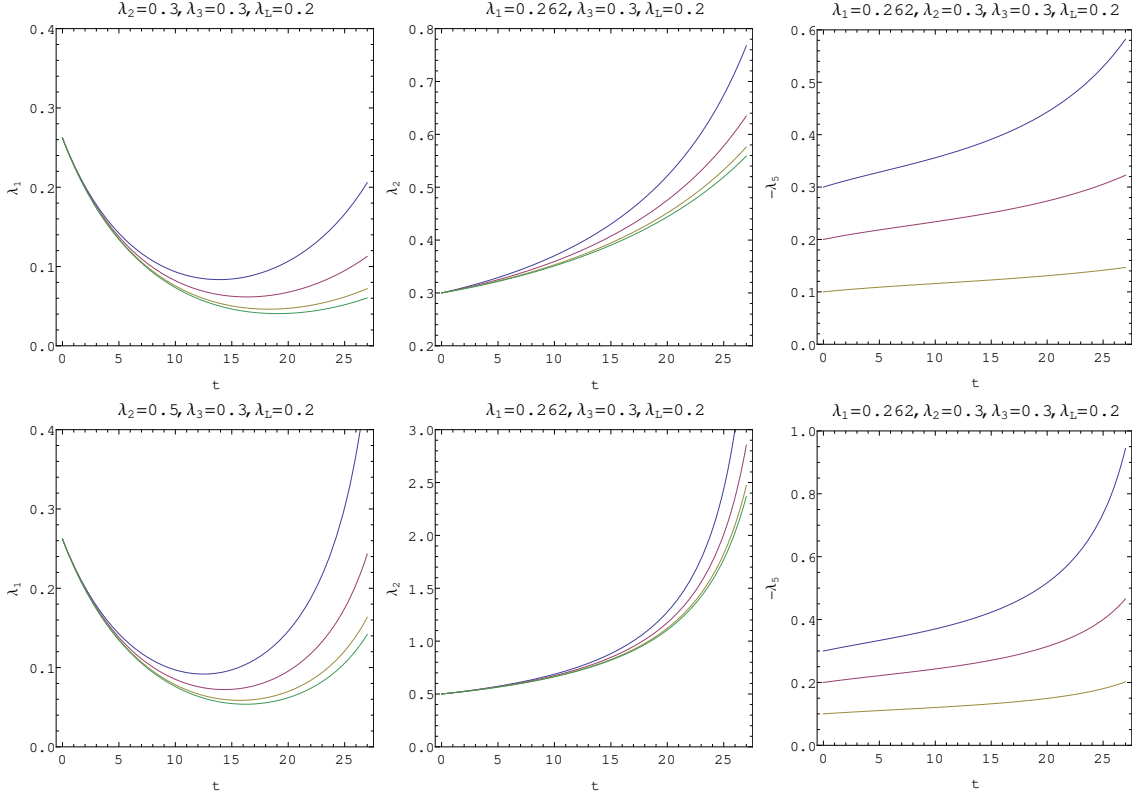}
\end{center}
\caption{(Upper row) Higgs quartic couplings versus $t=\log(\mu/m_W)$ for $\lambda_1=0.262$ (corresponding to $m_h=126\,{\rm GeV}$), $\lambda_2=0.3$ and $\lambda_5=-0.3$, $-0.2$, $-0.1$ and 0 at $\mu=m_t$ from top to bottom. At the unitarity scale, $t_U=\log(\mu_U/m_t)=26$.
In the lower row, the parameters are the same except $\lambda_2=0.5$.}
\label{fig:couplingsrunning}
\end{figure}

In Figure~\ref{fig:pureSMHiggsconst}, we depict the parameter space of the quartic couplings in the pure SM Higgs inflation consistent with the high scale constraints (\ref{pert}), (\ref{stable}) and (\ref{oneHcond1}), and the unbounded from below constraints (\ref{ufb}) at low energy.
Moreover, in Figure~\ref{fig:couplingsrunning}, some examples of the running quartic couplings consistent with the LHC limits on the Higgs mass are shown.
We note that the vacuum stability requires at least one of the couplings between the two Higgs bosons to be sizable in order for it to contribute to a positive running of the SM Higgs quartic coupling. There is more parameter space being compatible with the Higgs mass bound as compared to the multi-Higgs inflation.

We remark on unitarity in the pure single-Higgs inflation.
In the pure new Higgs inflation, as the extra quartic coupling can be unconstrained from below by experiments as in the inert doublet models, we can choose the extra quartic coupling $\lambda_2$ to be very small during inflation such that the non-minimal coupling $\xi_2$ is not large. On the other hand, in the pure SM Higgs inflation, we need to take a large non-minimal coupling $\xi_1$ because of the LEP bound on the Higgs mass. In this case as well as in the pure new Higgs inflation with large non-minimal couplings, unitarity violation is avoidable below the Planck scale, so the unitarization procedure along the line of Ref.~\cite{unitarization} may be taken.  Then,  while the scale of inflation becomes dependent on the unknown couplings of a new heavy scalar, the inflationary conditions
given by (\ref{oneHcond1}) can be unaltered~\cite{lebedev} as in the mixed Higgs inflation.

\subsection{Field-dependent cutoff and validity of the model}

As mentioned in the previous subsections, unitarity cutoff in the vacuum is given by ${\rm min}(\mpl/\xi_1,\mpl/\xi_2)$. So, from that point of view, unitarity problem should better be solved before the field values reach the unitarity cutoff \cite{unitarity}.
However, it has been noted~\cite{Bezrukov:2010jz} that the unitarity cutoff depends on the background field value in Higgs inflation.
In this section, we comment on the validity of our inflation model with the renormalizable potential for two Higgs doublets.

As shown in (\ref{potentialwomin}), at large field values, the Higgs interaction terms in the potential are Planck-suppressed.
Thus, we turn to the gauge interactions in a specific gauge (the analogue of unitary gauge in SM) by taking the solution for two Higgs doublets during inflation as in (\ref{higgsol}).
The gauge interactions in Einstein frame are
\begin{align}
\frac{{\cal L}_{\rm gauge}}{\sqrt{-g}} = & -\frac{1}{2\Omega^2}\bigg[g^2 h^2_1A^\mu A_\mu + \left( \partial_\mu\vartheta - g A_\mu \right)^2 h^2_2  \bigg]
\nonumber\\
= & -\frac{1}{2}\,\frac{1}{\xi_1+r^2 \xi_2} \Big(1-e^{-2\varphi/\sqrt{6}}\Big) \Big[ g^2A^\mu A_\mu +r^2 \left( \partial_\mu \vartheta - gA_\mu \right)^2  \Big] \, ,
\end{align}
where we have used the field redefinitions (\ref{newfield1}) and (\ref{newfield2}) at large field values satisfying $\xi_1 h^2_1+\xi_2 h^2_2\gg 1$.
Thus, due to the suppressed Higgs-gauge interactions,  we find that the gauge boson mass leads to the unitarity cutoff during inflation as
\begin{equation}
\Lambda_{UV}=\frac{\mpl}{\sqrt{\xi_1+r^2 \xi_2}} \, .
\end{equation}
On the other hand, the field values during inflation are $|h_1|\gg \Lambda_{UV}$, so higher order terms such as $c_n |\Phi_1|^{4+2n}/\Lambda^{2n}_{UV}$ might be problematic for slow-roll inflation.  For self-consistency of our model, we assume that the coefficients $c_n$ are set to small values at the scale $\Lambda_{UV}$ such that higher order terms are unimportant for inflation.  But, in a UV complete model along the line of a linear sigma model~\cite{unitarization}, higher order terms would be suppressed by the Planck scale, so the renormalizable potential of two Higgs doublet models could be valid at large Higgs field values without unitarity problem.
In the previous subsections, we have performed the RG analysis with renormalizable Higgs quartic couplings until the unitarity scale of the vacuum because we can use the SM RG equations.
The RG evolution above unitarity scale may depend on the UV completion \cite{newvstability}.


\section{Inert doublet models and low energy constraints}
\label{sec:lowenergy}

Phenomenology of the two Higgs doublet models depends on whether the extra Higgs doublet mixes with the SM Higgs doublet and how it interacts with the SM fermions.
There are two representative models with two Higgs doublets for minimal flavor violation, Type-I and Type-II 2HDMs. In Type-I models, the second Higgs doublet is odd under the $Z_2$ parity so that it does not couple to the SM fermions \cite{z2glashow}.  So, the second Higgs can couple to the SM fermions through the mixing with the SM Higgs.
In Type-II Higgs doublets models, one Higgs doublet couples to up-type quarks and the other Higgs doublet couples to down-type quarks and charged leptons, as in MSSM.

In this section, in order to consider the low energy constraints on the quartic couplings, we focus on the Type-I 2HDM with inert Higgs doublet where $\mu_3=0$ so $Z_2$ parity is an exact symmetry.
In this case, the second Higgs doublet does not obtain a VEV so the neutral scalar of the inert Higgs becomes a dark matter candidate \cite{inertdm}. Furthermore, the SM Higgs boson may decay invisibly into an inert Higgs pair when kinematically allowed. In this case, it is possible to constrain the quartic couplings mixing two Higgs doublets by the interplay between WMAP and LHC limits on the mass of the SM-like Higgs.

In the case where only the SM Higgs $h_1$ obtains a non-zero VEV,
we have $v^2=-2\mu^2_1/\lambda_1$ and need $\mu_3=0$. So, the $Z_2$ symmetry is respected even by dimensionful parameters. Then, the masses of the CP-even, CP-odd and the charged scalars are
\begin{align}
m^2_{h^0} =& \lambda_1 v^2 \, ,
\\
m^2_{H^0} =& \mu^2_2 +\frac{1}{2}(\lambda_L+\lambda_5) v^2 \, ,
\\
m^2_{A^0} =& \mu^2_2 + \frac{1}{2}(\lambda_L-\lambda_5) v^2 \, ,
\\
m^2_{H^\pm} =& \mu^2_2 +\frac{1}{2}\lambda_3 v^2 \, .
\end{align}
In the inert Higgs models in which the second Higgs doublet does not couple to the SM fermions, the lightest neutral scalar of the second Higgs doublet becomes a dark matter candidate. The condition for the Higgs dark matter is then $\lambda_L-\lambda_3-|\lambda_5|<0$. We take $H^0$ to be dark matter for $\lambda_5<0$.

\subsection{Low energy constraints}

We first enumerate the relevant low energy constraints on the inert doublet models, including accelerator bounds, electroweak precision data, dark matter constraints and the invisible Higgs decay at the LHC. In our numerical analysis, we have adopted the micrOMEGAs code~\cite{micromegas} to our model in order to compute the thermal WIMP relic abundance and the direct detection cross section.

\subsubsection{Accelerator bounds}

The mass of the charged scalar, $m_{H^\pm}$, is constrained to be larger than $70-90$ GeV \cite{limitMc} and $m_{H^0}+m_{A^0}$ must be larger than $m_Z$ to be compatible with $Z^0$-width measurements. The regions with $m_{H^0}<80\,{\rm GeV}$, $m_{A^0}<100\,{\rm GeV}$ and $m_{A^0}-m_{H^0}>8\,{\rm GeV}$ in the parameter space $(m_{H^0},m_{A^0})$ are excluded by LEP II data~\cite{limitMH}.

\subsubsection{Electroweak precision data}

An important constraint on the inert doublet model comes from electroweak precision data such as $S$ and $T$. The inert Higgs doublet gives a small contribution to $S$. Its contribution to $T$ is given by
\begin{equation}
\Delta T=\frac{1}{16\pi^2\alpha v^2} \Big[F\left(m_{H^\pm},m_{A^0}\right)+F\left(m_{H^\pm},m_{H^0}\right)-F\left(m_{A^0},m_{H^0}\right) \Big] \, ,
\end{equation}
where
\begin{equation}
F(m_1,m_2)\equiv\frac{1}{2}\left(m^2_1+m^2_2\right) - \frac{m^2_1 m^2_2}{m^2_1-m^2_2}\,\log \left( \frac{m^2_1}{m^2_2} \right) \, .
\end{equation}
On the other hand, the SM Higgs affects $T$ by
\begin{equation}
T_h\approx -\frac{3}{8\pi\cos^2\theta_W}\,\log \left( \frac{m_h}{m_Z} \right) \, .
\end{equation}
The electroweak precision constraint is
\begin{equation}
-0.1< \Delta T +T_h < 0.2 \, .
\end{equation}

\subsubsection{Dark matter constraints}

The extra neutral Higgs boson can be a dark matter candidate.
There are two kinds of interactions responsible for dark matter annihilations: one is the standard gauge interactions and the other is the extra couplings to the SM Higgs boson from the scalar potential. When the mass of dark matter satisfies $m_{H^0}<m_W$, annihilations through quartic couplings and/or co-annihilations through gauge interactions into the SM fermion-antifermion pair are dominant.
On the other hand, for $m_{H^0}>m_W$ and/or $m_{H^0}>m_h$, we need to consider additional channels including a pair of weak gauge bosons and/or the SM Higgs bosons in the final states.

From the Higgs quartic couplings, we obtain the annihilation cross section times velocity as
\begin{equation}
\langle\sigma_{f{\bar f}}v\rangle = \frac{1}{4\pi}\,(\lambda_L+\lambda_5 )^2m^2_f\, \frac{\left(1-m^2_f/m^2_{H^0}\right)^{3/2}}{\left(4m^2_{H^0}-m^2_h\right)^2} \, ,
\end{equation}
where $m_f$ is the SM fermion mass.
Then, from $\langle\sigma v\rangle=a+b v^2$, the relic density is given by
\begin{equation}
\Omega_{H^0}h^2=\frac{2.09\times 10^8\,{\rm GeV}^{-1}}{\mpl\sqrt{g_{*s}(x_F)}\,\left(a/x_F+3b/x^2_F\right)} \, ,
\end{equation}
where the freeze-out temperature gives $x_F=m_{H^0}/T_F\approx20$
and $g_{*s}(x_F)=45.75$.
In the case of multi-Higgs inflation, for a small $\lambda_5$, $H^0$ and $A^0$ masses are almost degenerate so the co-annihilation channels can be also important. Thus, due to a substantial number of $A^0$ at the time of freeze-out, the co-annihilation with $Z$ boson exchange can be important.
If $\lambda_5$ is sizable enough such that the splitting between $m_{H^0}$ and $m_{A^0}$ is larger than the freeze-out temperature $T_F$, the co-annihilation channel is suppressed so that the annihilation through the Higgs quartic couplings becomes the main channel for $m_{H^0}<m_W$. However,
as $m_{H^0}$ gets close to and go beyond $m_W$, additional channels including weak gauge bosons open up, becoming dominant in determining the relic density. We note that the relic density should be $0.094<\Omega_{H^0} h^2<0.136$ to be consistent with WMAP~\cite{Komatsu:2010fb} after taking into account theoretical and experimental systematic uncertainties. But we also show later the parameter space leading to $\Omega_{H^0}<0.136$ , having in mind that if the Higgs dark matter is not enough for giving the total dark matter relic abundance, other dark matter candidates such as the axion can constitute the rest.

For direct detection of dark matter, there are two leading diagrams to the spin-independent processes, $H^0 q\rightarrow A^0q$ with $Z$ boson exchange and $H^0 q\rightarrow H^0 q$ with the SM Higgs exchange. Experiments have reached the sensitivity to exclude the $Z$ boson exchange so we need $|m_{H^0}-m_{A^0}|\approx |\lambda_5|v^2/(2m_{H^0})\gtrsim 100\, {\rm keV}$ to forbid the first process kinematically.
For $m_{H^0}\sim 100$ GeV, we need $\lambda_5\gtrsim 10^{-6}$. If $\lambda_5$ is too small, a large co-annihilation cross section for dark matter would prevent us from obtaining a correct relic density in any case.

From the scattering process with the SM Higgs exchange, the spin-independent cross section for dark matter with nuclei is at tree level given by
\begin{equation}
\sigma^\text{(SI)}_{H^0\text{-}N}=\frac{(\lambda_L+\lambda_5)^2}{4\pi m^4_h}\,\frac{m^4_N f^2_N}{(m_{H^0}+m_N)^2} \, ,
\end{equation}
where $m_N$ is the nucleon mass and  $f_N\sim 0.3$ parametrizes the Higgs-nucleon coupling. There exist different estimations of the Higgs-nucleon coupling: the lattice result $f_N=0.326$~\cite{fNvalue1} and the MILC results with the minimal value $f_N=0.260$ and the maximal value $f_N=0.629$~\cite{fNvalue2}.
Thus, when we constrain the parameter space of our model by direct detection at XENON100~\cite{XENON}, we take into account those uncertainties and take the limit on the spin-independent cross section to be $\sigma^\text{(SI)}<5\times 10^{-8}\,{\rm pb}$ or $\sigma^\text{(SI)}<10^{-8}\,{\rm pb}$.
When the Higgs quartic couplings dominate the annihilation,  they also determine the cross section and are constrained by direct detection experiments.

\subsubsection{Invisible Higgs decay at the LHC}

If the SM Higgs mass satisfies $m_h> 2m_{H^0}$, it can decay into a dark matter pair so the branching fractions of the SM Higgs decay are changed.
The invisible decay rate is given by
\begin{equation}
\Gamma^{\rm inv}_{h\rightarrow H^0H^0}= \frac{(\lambda_L+\lambda_5)^2 v^2}{4\pi m_h}\sqrt{1-\frac{4m^2_{H^0}}{m^2_h}} \, .
\end{equation}
We note that the same coupling $\lambda_L+\lambda_5$ could determine not only dark matter relic abundance and direct detection but also invisible decay rate. Thus, the dark matter constraints could lead to the bound on the invisible decay rate and vice versa~\cite{lebedev,invisibleHiggs}.

\begin{figure}
\begin{center}
 \includegraphics[width=7.5cm]{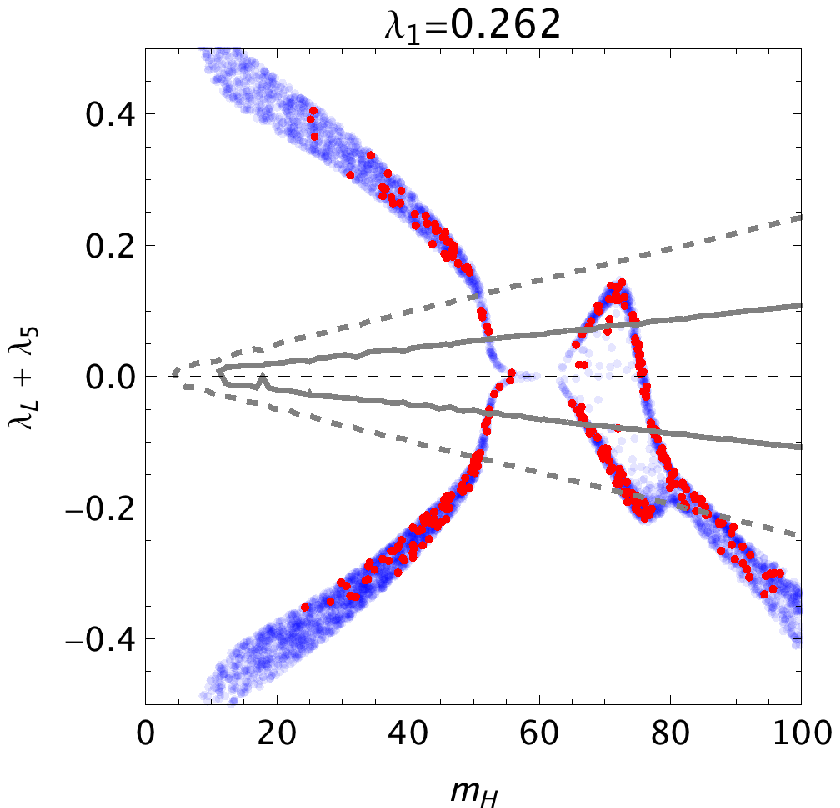}
 \includegraphics[width=7.5cm]{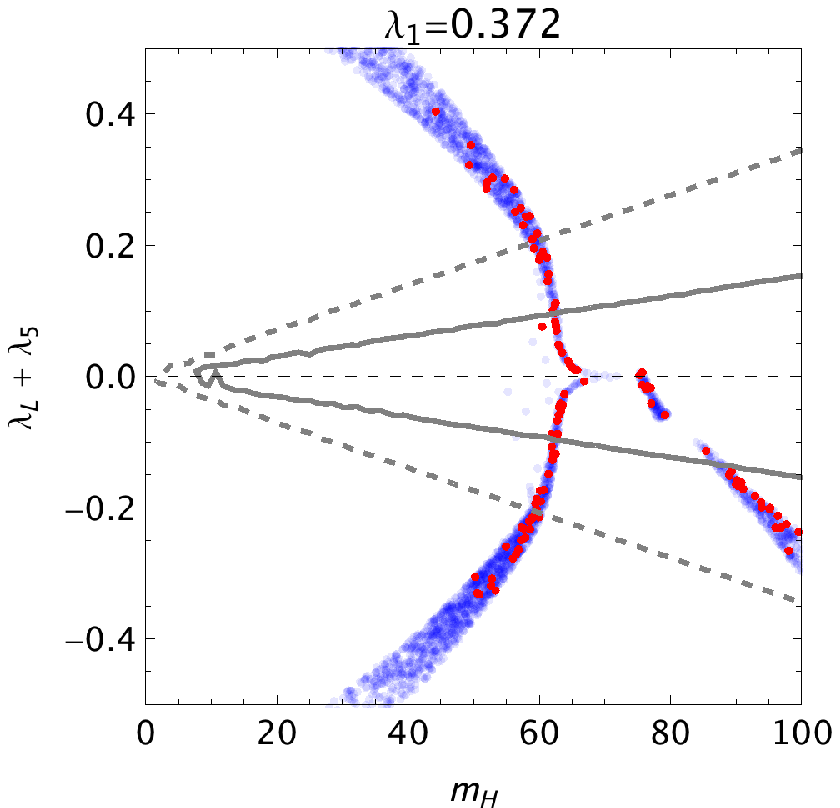}
\end{center}
\caption{(Blue) parameter space $(m_{H^0},\lambda_L+\lambda_5)$ consistent with collider/electroweak precision constraints, $0.094<\Omega_{\rm DM}h^2<0.136$ and direct detection: (dotted line) $\sigma^\text{(SI)}<5\times 10^{-8}{\rm pb}$ and (solid line) $\sigma^\text{(SI)}< 10^{-8}{\rm pb}$. Further constrained parameter space allowed by the pure SM Higgs inflation is shown in red. In the left (right) panel, we set $\lambda_1=0.262$ ($\lambda_1=0.372$), which corresponds to $m_h=126$ GeV ($m_h=150$ GeV).}
\label{fig:mHvslL+l5}
\end{figure}

\subsection{Implications from multi-Higgs inflation}

In the multi-Higgs inflation, we required the quartic coupling $\lambda_5$ to be very small during inflation. The other $U(1)_H$ breaking quartic couplings $\lambda_6$ and $\lambda_7$ must be equally small during inflation. $\lambda_5$ remains small under the running effects from the inflation scale to low energy because of the approximate $U(1)_H$ symmetry. Since $\lambda_5\lesssim 10^{-7}$ in most of the parameter space, a light dark matter lighter than 100 GeV is not possible because of too large co-annihilation, and is not compatible with direct detection experiments. However, when dark matter and pseudo-scalar Higgs are as heavy as $600\,{\rm GeV}$, it is possible to accommodate the dark matter relic density being compatible with direct detection. Because of a small splitting between $H^0$ and $A^0$ masses, electroweak precision constraints are satisfied even for the heavy extra Higgs bosons and other collider limits are satisfied as well.

\subsection{Implications from single-Higgs inflation}

In the single-Higgs inflation, $\lambda_5$ is sizable, reducing the co-annihilation channel for dark matter. Thus, a light dark matter lighter than 100 GeV is possible. In the left panel of  Figure~\ref{fig:mHvslL+l5}, we show the parameter space $(m_{H^0},\lambda_L+\lambda_5)$  for $m_h=126$ GeV, which is consistent with collider/electroweak precision data and the WMAP band on the relic density.

Focusing on the pure SM Higgs inflation, we find that the inflationary conditions restrict the modulus of the dark matter coupling to a small value less than 0.4. Direct detection constraint is stronger than the inflationary constraints, requiring the modulus of the dark matter coupling to be smaller than $0.1-0.2$. When dark matter mass is between 65 GeV and 80 GeV, the dark matter relic density depends on the sign of $\lambda_L+\lambda_5$ because the cancellation in the annihilation amplitude for $H^0H^0\to WW^{(*)}, ZZ^{(*)}$ can happen. The case with a higher SM Higgs mass $m_h=150$ GeV is as shown in the right panel of Figure~\ref{fig:mHvslL+l5} for comparison. In both cases, the invisible Higgs decay into a dark matter pair can be sizable such that the recent LHC limit on the heavier Higgs mass would not be applied and we need more sensitivity to discover the Higgs at the LHC~\cite{invisibleHiggs}.

\begin{figure}
\begin{center}
\includegraphics[width=15cm]{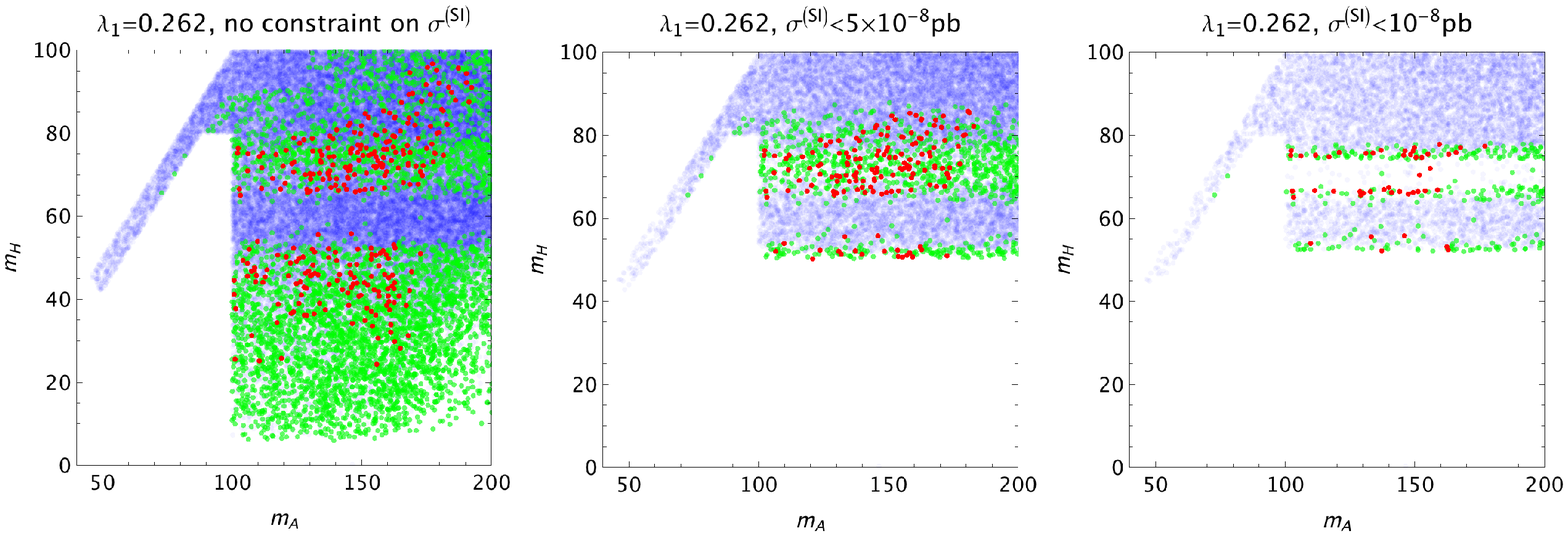}
\\
\includegraphics[width=15cm]{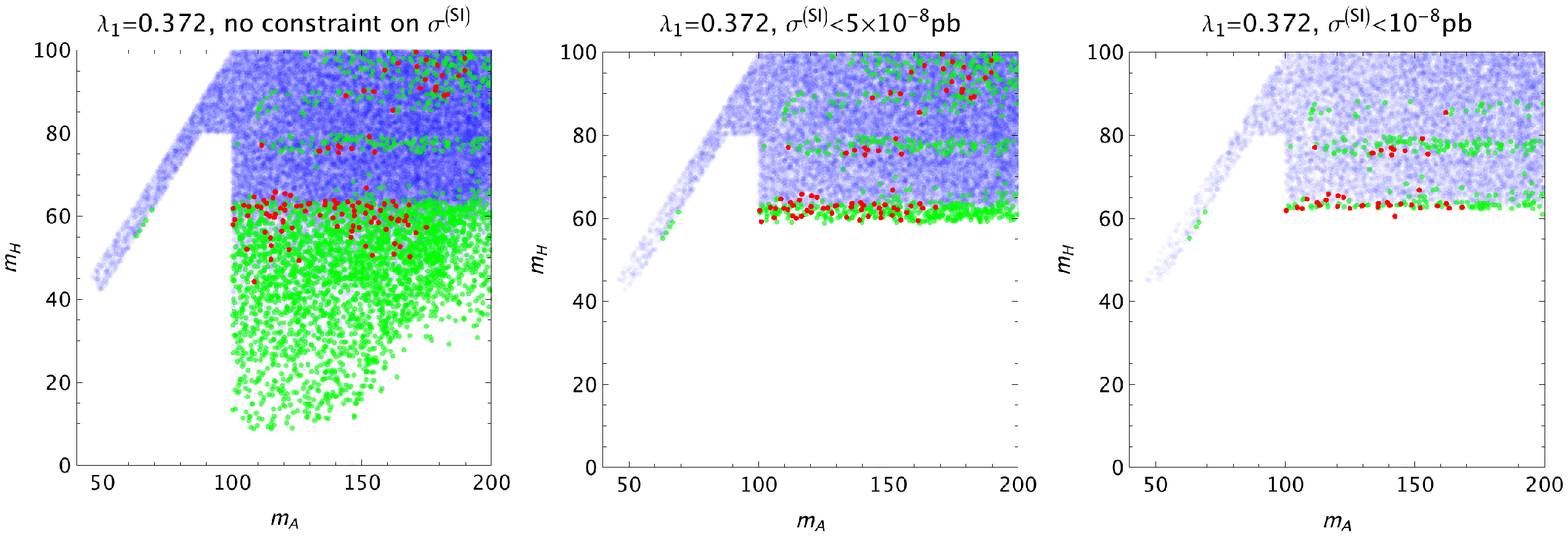}
\end{center}
\caption{Left: parameter region $(m_{A^0},m_{H^0})$ consistent with collider/electroweak precision constraints and dark matter relic density.
Middle: left $+$ direct detection with $\sigma^\text{(SI)}<5\times 10^{-8}{\rm pb}$.
Right: left $+$ direct detection with $\sigma^\text{(SI)}< 10^{-8}{\rm pb}$.
In each panel, the parameter space satisfying $(0.094<) \Omega_{\rm DM}h^2 < 0.136$ is in blue (green) and the region allowed by the pure SM Higgs inflation in addition to $0.094< \Omega_{\rm DM}h^2 < 0.136$ is in red. In the upper (lower) row, we set $\lambda_1=0.262$ ($\lambda_1=0.372$), which corresponds to $m_h=126$ GeV ($m_h=150$ GeV).}
\label{fig:mAvsmH}
\end{figure}

\begin{figure}
\begin{center}
\includegraphics[width=15cm]{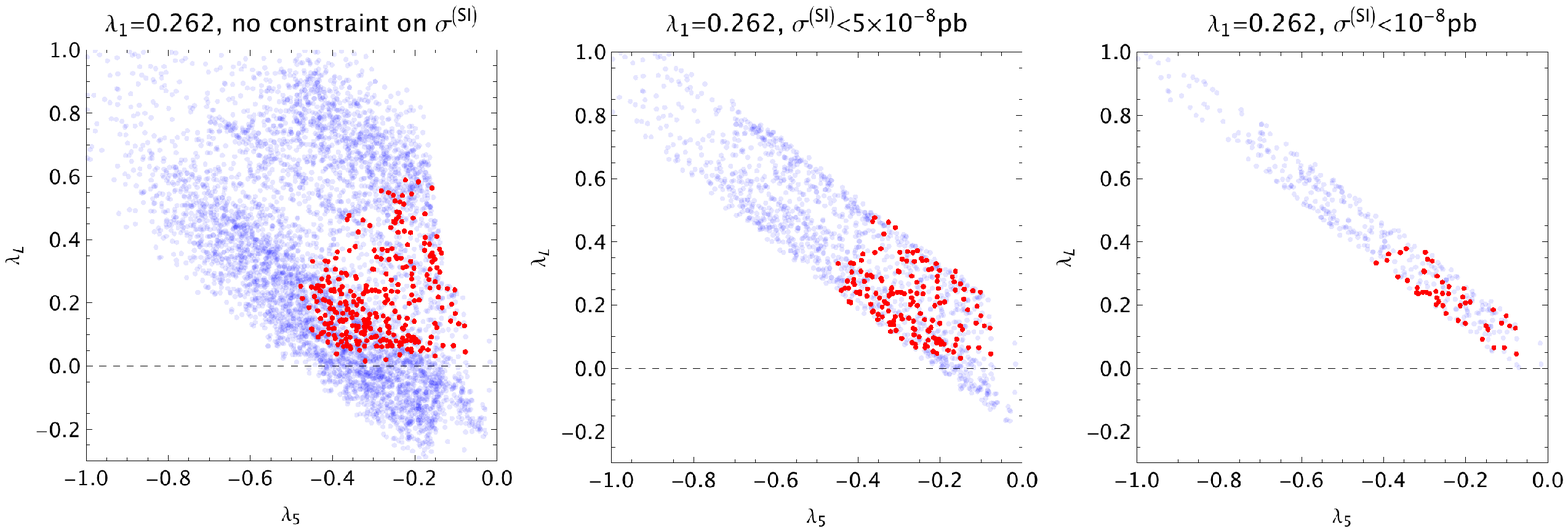}
\\
\includegraphics[width=15cm]{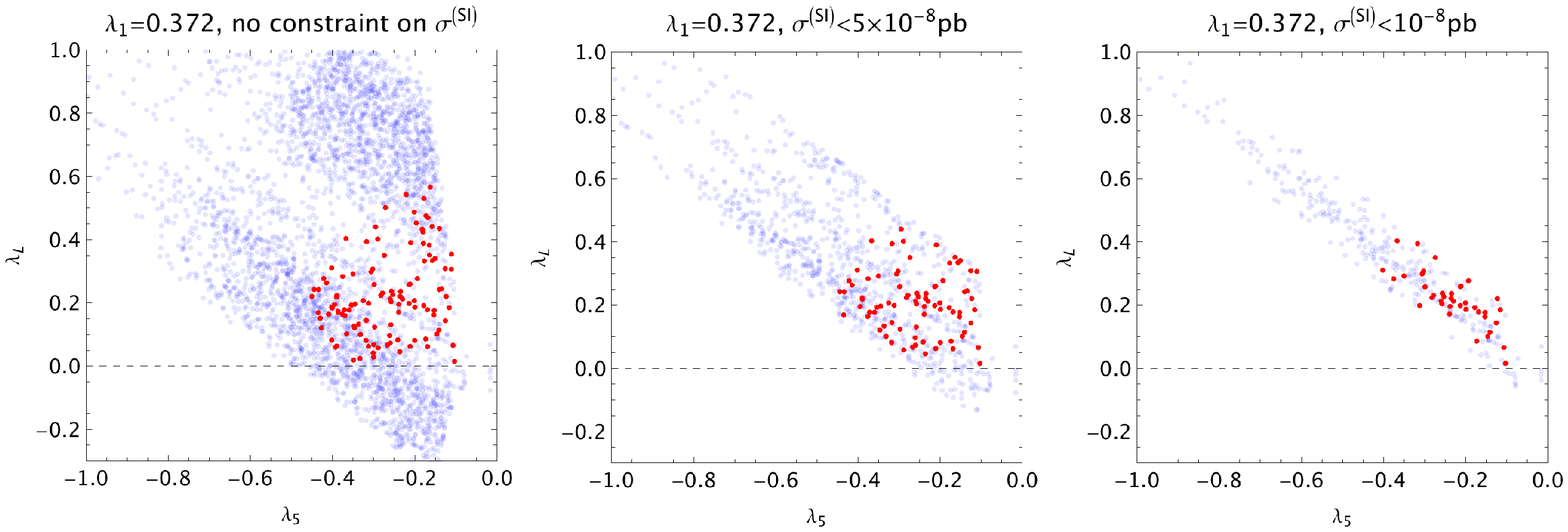}
\end{center}
\caption{Left: parameter region $(\lambda_5,\lambda_L)$ consistent with collider/electroweak precision constraints and $0.094<\Omega h^2<0.136$.
Middle: left $+$ direct detection with $\sigma^\text{(SI)}<5\times 10^{-8}{\rm pb}$.
Right: left $+$ direct detection with $\sigma^\text{(SI)}< 10^{-8}{\rm pb}$.
In each panel, the parameter space  allowed by the pure SM Higgs inflation in addition is in red.
In the upper (lower) row, we set $\lambda_1=0.262$ ($\lambda_1=0.372$), which corresponds to $m_h=126$ GeV ($m_h=150$ GeV).
}
\label{fig:l5vslL}
\end{figure}

In Figure~\ref{fig:mAvsmH}, the parameter space $(m_{A^0},m_{H^0})$ is shown both for $m_h=126\,{\rm GeV}$ and $m_h=150\,{\rm GeV}$, being consistent with dark matter relic density, collider/electroweak precision and direct detection.
We find that the inflationary conditions eliminate a significant fraction of the parameter space such that the light dark matter region is $50\,{\rm GeV}<m_{H^0}<80\,{\rm GeV}$  and $100\,{\rm GeV}<m_{A^0}<180\,{\rm GeV}$ for $\sigma^\text{(SI)}<10^{-8}\,{\rm pb}$.
In Figure~\ref{fig:l5vslL}, we show how the inflationary conditions influence the parameter space $(\lambda_5,\lambda_L)$ at low energy.
For $\sigma^\text{(SI)}<10^{-8}\,{\rm pb}$, the allowed parameter space becomes $0<\lambda_L<0.4$ and $-0.5<\lambda_5<-0.05$.
Note that for the heavier SM Higgs, there is a smaller parameter space satisfying the direct detection and the inflationary condition~(\ref{oneHcond1}), but the distinction between the light and heavy SM Higgs is not significant given a smaller upper limit on $\sigma^{\rm (SI)}$.

\section{Conclusion}
\label{sec:conc}

We have considered the inflationary scenario in two Higgs doublet models with non-minimal gravity couplings and discussed the interplay between the inflationary conditions and the low energy experimental constraints in restricting the parameter space of the extra quartic couplings.
For the phenomenology of two Higgs doublet models, we have taken the inert doublet models. The extra Higgs quartic couplings contribute to the positive running of the SM Higgs quartic coupling such that the vacuum stability is guaranteed until the unitarity scale in the extended Higgs inflation. Therefore, a sizable coupling between the SM Higgs and extra Higgs bosons is necessary.

Depending on whether the global $U(1)_H$ symmetry in the extended Higgs potential is approximate or not, there are two possibilities of the slow-roll inflation: one is the multi-Higgs inflation in the limit of an approximate $U(1)_H$ symmetry, and the other is the single-Higgs inflation in the case that $U(1)_H$ is broken to a $Z_2$ parity.
In the multi-Higgs inflation, large non-Gaussianity is possible for appropriate initial conditions on the pseudo-scalar Higgs during inflation and a negligible $U(1)_H$-breaking coupling $\lambda_5$ implies that the dark matter Higgs boson must be as heavy as 600 GeV.

In the single-Higgs inflation, there are three possibilities depending on the inflaton direction along the CP-even neutral Higgs bosons: the mixed Higgs inflation, the pure SM Higgs inflation and the pure new Higgs inflation.
In this case, a sizable $\lambda_5$ coupling allows for light dark matter below 100 GeV by suppressing the co-annihilation of dark matter. The stability condition for the single-Higgs inflation to occur gives rise to the additional conditions on the Higgs quartic couplings at the inflationary scale in addition to the usual perturbativity and vacuum stability conditions.
We have found that the inflationary conditions reduce the parameter space of the mass and couplings of dark matter at a level that the detection at future dark matter experiments is possible and/or a sizable invisible Higgs decay influences the current Higgs search at the LHC. It will be interesting to look at the consequences of the inflationary conditions for the other two Higgs doublet models than the inert doublet models. We leave the study on this in a future work.

\acknowledgments
We would like to thank Myeonghun Park, Alexander Pukhov and Hwa Sung Cheon for helps with running the micrOMEGAs code, and Michael Trott for the discussion on two Higgs doublet models.
JG and HML are partially supported by Korean-CERN fellowship. SKK is in part supported by the National Research Foundation of Korea
(NRF) grant funded by the Korea government of the Ministry of Education, Science and
Technology (MEST) (No. 2011-0027227).

\appendix

\section{Minimization of the potential during inflation}
\label{app:stabilization}{\small

We consider the minimization of the potential (\ref{rpot}) with respect to $r$.
The extremum condition for $r$ is
\begin{equation}
0=\frac{\partial V}{\partial r}=\frac{r(ar^2-b)}{2\left(\xi_1+\xi_2 r^2\right)^3} \, ,
\end{equation}
where
$a\equiv \lambda_2\xi_1-\lambda_L\xi_2$ and $b\equiv \lambda_1\xi_2-\lambda_L\xi_1$.
Thus, there are three extrema at $r^2_0=0$, $\infty$ and $b/a$.
The double derivative of the potential is
\begin{equation}
\frac{\partial^2V}{\partial r^2}\bigg|_{r^2=r^2_0}
= \frac{\left(3a r^2_0-b\right)\left(\xi_1+\xi_2 r^2_0\right) - 6\xi_2 r^2_0\left(ar^2_0-b\right)}{2\left(\xi_1+\xi_2 r^2_0\right)^4} \, .
\end{equation}
The conditions for a stable minimum and the vacuum energy $V_0$ for each case are given in the following table:
\begin{table}[h!]\small
\begin{center}
\begin{tabular}{c||c|c}
 & stable minimum condition & $V_0$
\\\hline\hline
$r_0^2=0$ & $a>0$ and $b<0$ & $\dfrac{\lambda_1}{8\xi_1^2}$
\\
$r_0^2=\infty$ & $a<0$ and $b>0$ & $\dfrac{\lambda_2}{8\xi_2^2}$
\\
$r_0^2=\dfrac{b}{a}$ & $a>0$ and $b>0$ & $\dfrac{\lambda_1\lambda_2-\lambda_L^2}{8\left(\lambda_1\xi^2_2+\lambda_2\xi^2_1-2\lambda_L\xi_1\xi_2\right)}$
\end{tabular}
\end{center}
\end{table}

For $r_0=0$ or $r_0=\infty$, one of the CP-even neutral Higgs bosons drives inflation but there is no potential for the pseudo-scalar Higgs. For $r^2_0=b/a$, for a positive vacuum energy, we need an additional condition, $\lambda_1\lambda_2-\lambda_L^2>0$.

\section{Kinetic mixing and decoupling of heavy state}
\label{app:kinmixing}{\small

In the presence of the kinetic mixing, it is important to identify the heavy state to obtain the effective theory for the light inflaton by integrating out the heavy mode consistently~\cite{integratingout}.
For simplicity, we first consider the effect of the kinetic mixing for $\xi_2=0$. In this case, the kinetic term in (\ref{largefieldaction}) becomes
\begin{equation}
\calL_{\rm kin}
\approx -\frac{1}{2}\left(1+\frac{1}{6\xi_1}\right)(\partial_\mu\varphi)^2-\frac{1}{2}e^{-2\varphi/\sqrt{6}}\,\Big[(\partial_\mu h_2)^2+h^2_2(\partial_\mu\vartheta)^2\Big] \, ,
\end{equation}
where $h_2=h_1r=r\,e^{\varphi/\sqrt{6}}/\sqrt{\xi_1}$.
Therefore, the kinetic mixing term is absent, though the kinetic terms for $h_2$ and $\vartheta$ are still non-canonical. But, during the slow-roll motion of the inflaton $\varphi$ at non-zero value, the coefficient of the kinetic terms for $h_2$ and $\vartheta$ are almost constant.
Then, the $r$-dependent part of the potential (\ref{rpot}) can be written as
\begin{equation}
V_{r\text{-dep}} \approx \frac{1}{8\xi_1}\Big(\lambda_1+\lambda_2 \xi^2_1 e^{-4\varphi/\sqrt{6}}h^4_2+2\lambda_L\xi_1 e^{-2\varphi/\sqrt{6}}h^2_2\Big) \, .
\end{equation}
From the minimization condition for $h_2$, we obtain $h^2_2=r^2_0e^{2\varphi/\sqrt{6}}/\xi_1$ with $r^2_0=-\lambda_L/\lambda_2$.
This minimum with positive vacuum energy exists only for $-\sqrt{\lambda_1\lambda_2}<\lambda_L<0$. Consequently, we obtain the mass of the heavy Higgs as
\begin{equation}
m^2_{h_2}=e^{2\varphi/\sqrt{6}}\,\frac{\partial^2 V}{\partial h^2_2}=-\frac{\lambda_L}{\xi_1} \, .
\end{equation}
Therefore, for $|\lambda_L|\lesssim {\cal O}(1)$, the mass of the heavy Higgs is of  $\calO(1/\sqrt{\xi_1})$. On the other hand, $H=\calO(\sqrt{\lambda_{\rm eff}}/\xi_1)$ with $\lambda_{\rm eff}=\big(\lambda_1-\lambda_L^2/\lambda_2\big)/2$. Thus, for $\xi_1\gtrsim 1$, the mass of the heavy Higgs is $m_{h_2}\gtrsim H$, so the heavy state quickly settles to a minimum.
Then, the remaining fields, $\varphi$ and $\vartheta$, play the role of the inflaton.

For a non-zero $\xi_2$, we could not find new variables to get rid of the kinetic mixing term for arbitrary field values. However, we can still prove that the heavy Higgs state can be safely integrated out as follows. Suppose that the field ratio $r$ has been fixed at a finite value as discussed in the previous section.
Then, from (\ref{largefieldaction}), the kinetic terms for $\varphi$ and the perturbation ${\bar r}$ (from $r=r_0+{\bar r}$) can be written as
${\cal L}_{\rm kin}=-K_{ij}\partial_\mu\phi^i\partial^\mu\phi^j/2$,
where the components of $K_{ij}$ are $K_{\varphi\varphi}\approx 1$, $K_{\varphi {\bar r}}=K_{{\bar r} \varphi}\equiv \alpha$ and $K_{{\bar r}{\bar r}}=\beta$. Then, the kinetic terms can be diagonalized
by choosing the new variables, $\varphi'=\cos\Theta \,\varphi+\sin\Theta \,{\bar r}$ and ${\bar r}'=-\sin\Theta\,\varphi+\cos\Theta \,{\bar r}$, with $\tan\Theta=2\alpha/\big[1-\beta+\sqrt{(1-\beta)^2+4\alpha^2}\big]$.
In this case, the eigenvalues of the kinetic terms are $\lambda_\pm=\big[1+\beta\pm \sqrt{(1-\beta)^2+4\alpha^2}\big]/2$.
On the other hand, the $r$-dependent part of the potential (\ref{rpot}) is expanded around the minimum as $V \approx V_0+A{\bar r}^2$ with  $A\equiv(\lambda_2\xi_1-\lambda_L\xi_2)/(\xi_2 r^2_0+\xi_1)^3>0$. So, the potential can be expanded in terms of new field variables as
\begin{equation}
V \approx V_0+A\Big(\sin\Theta\,\varphi+\cos\Theta\, {\bar r}'\Big)^2 \, .
\end{equation}
By rescaling the field variables with ${\widehat\varphi}=\sqrt{\lambda_+}\varphi$ and ${\widehat r}=\sqrt{\lambda_-}{\bar r}'$, the kinetic terms are canonically normalized, while the mass matrix for ${\widehat \varphi}$ and ${\widehat r}$ have the components, $M^2_{{\widehat\varphi}{\widehat\varphi}}=A\sin^2\Theta/\lambda_+$, $M^2_{{\widehat\varphi}{\widehat r}}=M^2_{{\widehat r}{\widehat\varphi}} =A\sin\Theta\cos\Theta/\sqrt{\lambda_+\lambda_-}$ and $M^2_{{\widehat r}{\hat r}}=A\cos^2\Theta/\lambda_-$. Therefore, we find that the mass eigenvalues are
\begin{align}
m^2_+ = & A \left( \frac{\sin^2\Theta}{\lambda_+}+\frac{\cos^2\Theta}{\lambda_-} \right) \, ,
\\
m^2_- = & 0 \, .
\end{align}
Then, the $\varphi$-dependent part of the potential in (\ref{largefieldaction}) gives rise to a slow-roll potential for the massless mode, but the changes to the minimum value of $r$ and the mass of the heavy state are exponentially suppressed. Consequently, for large non-minimal couplings, we obtain $A\sim \lambda_2/\xi^2_{\rm eff}$, $\Theta\sim 1/\xi_{\rm eff}\ll 1$, $\lambda_+\sim 1$ and $\lambda_-\sim r_0/\xi_{\rm eff}$, so the mass of the heavy state becomes $m^2_+\sim \lambda_2/(r_0\xi_{\rm eff})$, which is much larger than the Hubble parameter, $H^2\sim \lambda_{\rm eff}/\xi^2_{\rm eff}$.
Even for $\xi_{\rm eff}\sim 1$, the heavy state obtains mass of $\calO(H)$, so we can safely integrate out the heavy state.

\section{Renormalization group equations for the inert doublet model}
\label{app:RG}{\small

The RG equations for $p_i$ are defined by $\partial p_i/\partial t=\beta_{p_i}$
with $t=\log(\mu/m_W)$, where $\beta_{p_i}$ is the corresponding beta function.
In the inert doublet model, setting $\xi_3=\lambda_6=\lambda_7=0$, the beta functions of the remaining quartic couplings for two Higgs doublets are the following~\cite{sm2hd,RG},
\begin{align}
16\pi^2\beta_{\lambda_1} = & 12\lambda^2_1+4\lambda^2_3+4\lambda_3\lambda_4+2\lambda^2_4+2\lambda^2_5
+\frac{3}{4}\left[2g^4+\left(g^2+{g'}^2\right)^2\right]-12h^4_t-64\pi^2\lambda_1\gamma_1 \, ,
\\
16\pi^2\beta_{\lambda_2} = & 12\lambda^2_2+4\lambda^2_3+4\lambda_3\lambda_4+2\lambda^2_4+2\lambda^2_5
+\frac{3}{4}\left[2g^4+\left(g^2+{g'}^2\right)^2\right]-64\pi^2\lambda_2\gamma_2 \, ,
\\
16\pi^2\beta_{\lambda_3} = & 2(\lambda_1+\lambda_2)(3\lambda_3+\lambda_4)+4\lambda^2_3+2\lambda^2_4+2\lambda^2_5
+\frac{3}{4}\left[2g^4+\left(g^2-{g'}^2\right)^2\right]-32\pi^2\lambda_3 (\gamma_1+\gamma_2) \, ,
\\
16\pi^2\beta_{\lambda_4} = & 2\lambda_4(\lambda_1+\lambda_2+4\lambda_3+2\lambda_4)+8\lambda^2_5
+3g^2{g'}^2-32\pi^2\lambda_4 (\gamma_1+\gamma_2) \, ,
\\
16\pi^2 \beta_{\lambda_5} = & 2 \lambda_5(\lambda_1+\lambda_2+4\lambda_3+6\lambda_4)-32\pi^2\lambda_5 (\gamma_1+\gamma_2) \, ,
\end{align}
where the anomalous dimensions of the two Higgs doublets are
\begin{align}
\gamma_1 = & \frac{1}{64\pi^2}\left(9g^2+3{g'}^2-12h^2_t\right) \, ,
\\
\gamma_2 = & \frac{1}{64\pi^2}\left(9g^2+3{g'}^2\right) \, .
\end{align}
The beta functions of the Yukawa couplings and the gauge couplings are
\begin{align}
16\pi^2\beta_{h_t} = & h_t \left( \frac{9}{2}h^2_t-8g^2_3-\frac{9}{4}g^2-\frac{17}{12}{g'}^2 \right) \, ,
\\
16\pi^2\beta_{g'} = & 7{g'}^3 \, ,
\\
16\pi^2\beta_{g} = & -3g^3 \, ,
\\
16\pi^2\beta_{g_3} = & -7g^3_3 \, .
\end{align}

On the other hand, the beta functions for the mass parameters are~\cite{massrg}
\begin{align}
\label{massRG1}
16\pi^2 \beta_{\mu^2_1} = & 6\mu^2_1\lambda_1 +2\mu^2_2(2\lambda_3+\lambda_4) -32\pi^2 \gamma_1 \mu^2_1 \, ,
\\
\label{massRG2}
16\pi^2\beta_{\mu^2_2} = & 6\mu^2_2\lambda_2 +2\mu^2_1(2\lambda_3+\lambda_4) -32\pi^2 \gamma_2 \mu^2_2 \, .
\end{align}
The bare non-minimal coupling $\xi_o$ is written in terms of the renormalized one $\xi$ and the mass renormalization $Z_m$ as
$\xi_{oij}=(\xi_{kl}+ \delta_{kl}/6)Z^{kl}_{mij}-\delta_{ij}/6$. Thus, the RG equation for the non-minimal coupling is~\cite{buchbinder}
\begin{equation}
\frac{\partial}{\partial\log\mu}\,\xi_{ij}= \left(\xi_{kl}+\frac{1}{6}\delta_{kl}\right)\gamma^{kl}_{mij} \, ,
\end{equation}
where $\beta_{m^2_{ij}}=\gamma^{ab}_{mij}m^2_{ab}$.
Consequently, using (\ref{massRG1}) and (\ref{massRG2}), the beta functions of the non-minimal couplings for two Higgs doublets are
\begin{align}
\label{xi11}
16\pi^2\beta_{\xi_1} = & \left(\xi_1+\frac{1}{6}\right) \left(6\lambda_1-32\pi^2\gamma_1 \right) + \left(\xi_2+\frac{1}{6}\right)(4\lambda_3+2\lambda_4) \, ,
\\
\label{xi22}
16\pi^2\beta_{\xi_2} = & \left(\xi_2+\frac{1}{6}\right)(6\lambda_2-32\pi^2\gamma_2)+ \left(\xi_1+\frac{1}{6}\right)(4\lambda_3+2\lambda_4) \, .
\end{align}

\end{document}